\documentclass[aps,superscriptaddress,showpacs,floatfix]{revtex4}
\usepackage[latin9]{inputenc}
\usepackage{amsmath}
\usepackage{amssymb}
\usepackage{graphicx}
\usepackage{esint}
\usepackage{epstopdf}
\usepackage{braket}
\usepackage{hyperref}
\usepackage{color}
\usepackage[export]{adjustbox}

\makeatletter

\begin{document}
\unitlength = 1mm

\title{Quantum transport in the black-hole configuration of an atom condensate outcoupled through an optical lattice}

\author{J. R. M. de Nova}
\affiliation{Department of Physics, Technion-Israel Institute of Technology, Technion City, Haifa 32000, Israel}
\author{F. Sols}
\affiliation{Departamento de Física de Materiales, Universidad Complutense de
Madrid, E-28040 Madrid, Spain}
\author{I. Zapata}
\affiliation{Departamento de Física de Materiales, Universidad Complutense de
Madrid, E-28040 Madrid, Spain}

\begin{abstract}
The outcoupling of a Bose-Einstein condensate through an optical lattice provides an interesting scenario to study quantum transport phenomena or the analog Hawking effect as the system can reach a quasi-stationary black-hole configuration. We devote this work to characterize the quantum transport properties of quasi-particles on top of this black-hole configuration by computing the corresponding scattering matrix. We find that most of the features can be understood in terms of the usual Schr\"odinger scattering. In particular, a transmission band appears in the spectrum, with the normal-normal transmission dominating over the anomalous-normal one. We show that this picture still holds in a realistic experimental situation where the actual Gaussian envelope of the optical lattice is considered. A peaked resonant structure is displayed near the upper end of the transmission band, which suggests that the proposed setup is a good candidate to provide a clear signal of spontaneous Hawking radiation.
\end{abstract}

\pacs{03.75.Kk 04.62.+v 04.70.Dy \volumeyear{2012} \volumenumber{number}
\issuenumber{number} \eid{identifier} \startpage{1}
\endpage{}}

\date{\today}

\maketitle

\section{Introduction}

The spontaneous emission of radiation with a thermal Planck spectrum by the event horizon of a black hole was predicted by Hawking using a semiclassical model in which matter was quantized and gravitation was treated as classical \cite{Hawking1974,Hawking1975}. The detection of such intriguing effect still remains one of the main open questions in modern Physics. The small effective temperature of emission, even lower than the cosmic microwave background, contributes to make the potential observation extremely difficult. However, it was shown by Unruh \cite{Unruh1976,Unruh1981} that quantum fluids traversing a subsonic-supersonic interface, which plays the analog role of an event horizon, could also exhibit a similar effect at temperatures which, while still low, at least seem experimentally reachable. Since then, many proposals have been made to detect the analog of Hawking radiation in systems such as different as Fermi gases \cite{Giovanazzi2005}, ion rings \cite{Horstmann2010} or polaritons \cite{Nguyen2015}.

Among all the analogues, the specific case of Bose-Einstein condensates \cite{Garay2000} is one of the most promising setups to study analog effects in the laboratory due to their low temperature, the relative ease of handling, and the availability of a deep understanding of their quantum behavior. The role of the particle-antiparticle pair emitted at the horizon is here played by the spontaneous entangled emission of phonons into the subsonic and supersonic regions \cite{Leonhardt2003,Leonhardt2003a,Balbinot2008,Carusotto2008,Recati2009,Macher2009,Coutant2010,deNova2014,Finazzi2013,Busch2014,deNova2015}. Indeed, sonic event horizons have already been experimentally produced by accelerating a condensate \cite{Lahav2010}. In a similar configuration, the analogue of self-amplifying Hawking radiation (the so-called black-hole laser \cite{Finazzi2010}) has been also observed although the interpretation of the results of that experiment is still under some controversy \cite{Tettamanti2016,Wang2016,Steinhauer2017}. Finally, the observation was recently reported of the spontaneous emission of Hawking radiation by a sonic horizon through the detection of its entanglement properties \cite{Steinhauer2016}, which constitutes the first experimental evidence of the (analog) Hawking effect.

An alternative route to the production of analog black holes is the quasi-stationary leaking of a large condensate reservoir \cite{Zapata2011,Larre2012,deNova2015} or the launching of a condensate against a localized obstacle \cite{Kamchatnov2012,Gerace2012}, with an outgoing flux which is dilute enough to be supersonic. In particular, Ref. \cite{deNova2015} presented a detailed numerical study of a realistic experimental protocol in which a finite-size condensate is confined on one side by an optical lattice, whose amplitude is gradually lowered until a quasi-stationary black-hole configuration is reached. The goal of this work is to study the corresponding quasi-particle transport properties on top of the obtained quasi-stationary mean-field solution by computing numerically the scattering coefficients, devoting special attention to the anomalous elements that characterize the spontaneous Hawking emission.

Besides the clear gravitational analogue, the study of such quasi-stationary scenarios is of general interest for the investigation of atom quantum transport \cite{Bloch1999,Andersson2002,Paul2007,Guerin2006,PhysRevA.80.041605,PhysRevA.84.043618,Brantut2012,Brantut2013}, within the emergent field of atomtronics \cite{Seaman2007,Labouvie2015}. For instance, the considered setup can be used to provide a quasi-stationary supersonic current, with a well controlled atom velocity.

This paper is arranged as follows. Section \ref{sec:themodel} briefly reviews the main properties of the mean-field black-hole configuration here considered. The numerical results for the computation of the scattering matrix on top of the described mean-field configuration are shown in Sec. \ref{sec:quasistationaryscattering}, which constitute the central contribution of this work. Conclusions and perspectives are outlined in Sec. \ref{sec:conclusions}. The general formalism of gravitational analogues in Bose-Einstein condensates is reviewed in Appendix \ref{app:generalformalism}, while the numerical methods for computing the scattering matrix are explained in Appendix \ref{app:numerical}.

\section{Black hole in an outcoupled Bose-Einstein condensate through an optical lattice} \label{sec:themodel}

We review in this section the physical system under consideration, which was thoroughly studied in Ref. \cite{deNova2014a}, where the reader is referred for more details. A detailed discussion of gravitational analogues and the mean-field formalism for Bose-Einstein condensates is presented in Appendix \ref{app:generalformalism}.

We consider the outcoupling of a one-dimensional (1D) Bose-Einstein condensate, initially confined at $t=0$ on one side by an optical lattice whose amplitude is gradually lowered from $V_{0}$ to some asymptotic amplitude $V_{\infty}$, so that a quasi-stationary black-hole configuration can be reached. Specifically, working in the 1D mean-field regime \cite{Leboeuf2001,Menotti2002}, the time evolution of the system for $t>0$ is described by the 1D time-dependent GP equation:
\begin{equation}\label{eq:TDGPOL}
i\hbar\frac{\partial\Psi(x,t)}{\partial t}=\left[-\frac{\hbar^{2}}{2m}\partial_{x}^{2}+V(x,t)+g|\Psi(x,t)|^{2}\right]\Psi(x,t)\, ,
\end{equation}
with $V(x,t)$ the time-dependent optical lattice potential, which is chosen to be of the form
\begin{eqnarray}\label{eq:TDPotential}
\nonumber V(x,t) & = & V(t)f(x) \\
V(t) & = & V_{\infty}+(V_{0}-V_{\infty})e^{-t/\tau} \, ,
\end{eqnarray}
$f(x)$ being some dimensionless function that describes the shape of the lattice (its form depends on the considered case, see Secs. \ref{subsec:ideal}, \ref{subsec:Gaussian-shaped}). The condensate density is nonzero only for $x>0$ because a sufficiently high confining barrier is placed at $x=0$ for all times, taken into account through a hard-wall boundary condition $\Psi(0,t)=0$. Before the lowering of the lattice amplitude, $t\leq 0$, the system is confined at equilibrium by the same optical lattice with constant amplitude $V(0)=V_{0}$. Hence, as initial condition for the wave function we take the stationary GP wave function, $\Psi(x,0)=\Psi_0(x)$, given by the solution of the corresponding time-independent GP equation
\begin{eqnarray}\label{eq:GPinitial}
\left[-\frac{\hbar^{2}}{2m}\partial_{x}^{2}+V(x,0)+g|\Psi_0(x)|^{2}-\mu_{0}\right]\Psi_0(x) & = & 0 \, ,
\end{eqnarray}
that describes the initial condensate of $N$ atoms at equilibrium, with the initial chemical potential $\mu_0$ determined by the normalization condition $\int\mathrm{d}x~|\Psi_0(x)|^{2}=N$.
This chemical potential also defines some relevant physical scales: density $n_0\equiv \mu_{0}/g$, length $\xi_0\equiv \sqrt{\hbar^{2}/mgn_{0}}$, velocity $c_{0}\equiv \sqrt{gn_{0}/m}$, and time $t_0\equiv \hbar/\mu_{0}$; all these quantities will be useful to characterize the different magnitudes of the problem. Note that $\xi_0,c_0$ are the corresponding healing length and sound speed, respectively.

The qualitative picture of the evolution of the system is the following: initially, at $t=0$, we have a confined 1D condensate occupying the region $0<x\lesssim L$, so $n_{0}\sim N/L$. The optical lattice has a typical size $L_{\rm lat}$ and is present in the region $L \lesssim x \lesssim L+L_{\rm lat}$, with initial amplitude sufficiently large ($V_{0}\gg\mu_{0}$) to provide a strong confinement. The remaining free (without confining potential) region $L+L_{\rm lat}\lesssim x$ will be denoted as the supersonic or downstream region, since we expect the escaping flow there to be sufficiently dilute to be supersonic. Although in theory the supersonic region is infinite, in numerical computations it must be finite; for that purpose, and also to reduce spurious numerical artifacts, absorbing boundary conditions (ABC) are implemented in the downstream region \cite{deNova2014a}.

\begin{figure}[tb!]
\includegraphics[width=1\columnwidth]{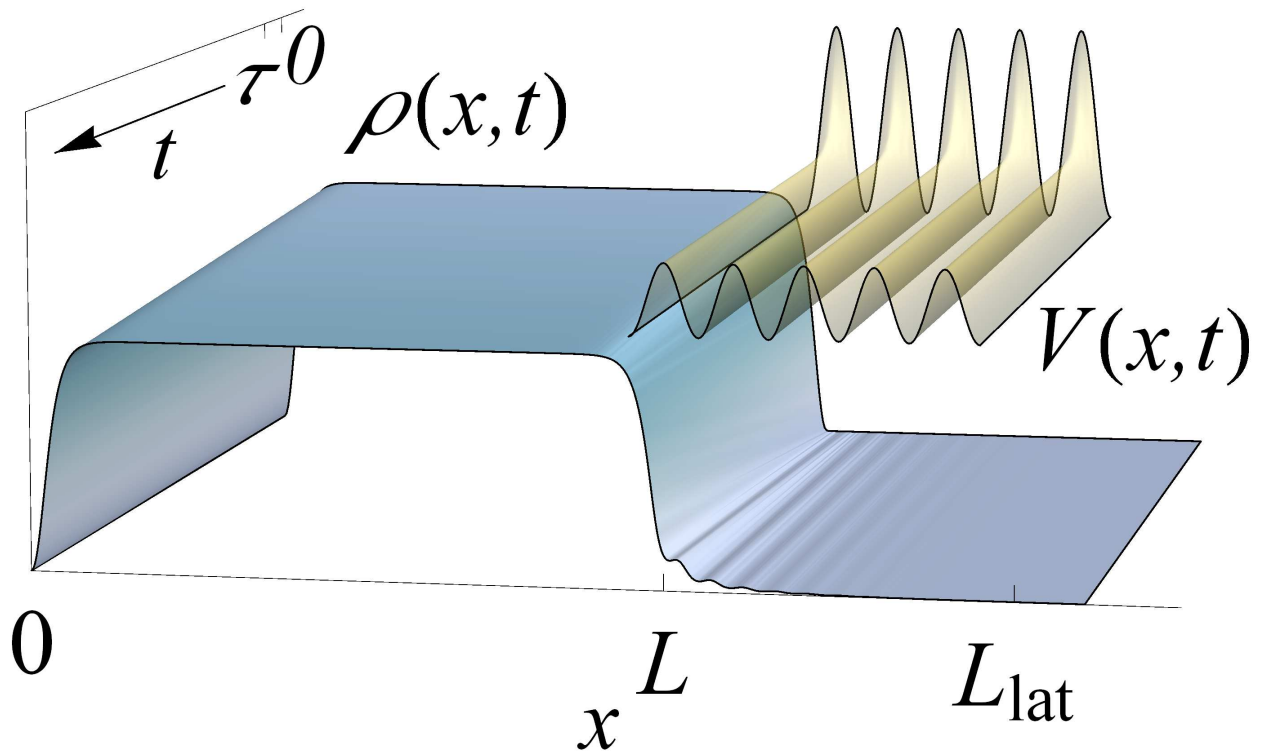}
\caption{Schematic representation of the evolution of the outcoupling condensate. A hard-wall boundary condition is assumed at $x=0$ and
an optical lattice lies in the region $L \lesssim x \lesssim L+L_{\rm lat}$ (where $L+L_{\rm lat}$ is labeled as $L_{\rm lat}$ in the plot) with a time-dependent amplitude such that the potential
$V(x,t)$ (represented by the semi-transparent yellow surface over the $x-t$ plane) evolves from strongly to moderately confining. The resulting time-dependent density profile $n(x,t)$ [labeled as $\rho(x,t)$ in the plot] is represented by the grey-blue surface.
The vertical axis represents the density. The surface $V(x,t)$ is uplifted to provide a better vision of $n(x,t)$. The trend towards a long-time quasi-stationary flow regime can be qualitatively observed. Figure taken from Ref. \cite{deNova2014a}.}
\label{fig:Scheme}
\end{figure}

Then, for $t>0$, the time-dependent amplitude $V(t)$ of the barrier evolves from $V_{0}\gg\mu_{0}$ to $V_{\infty}\gtrsim\mu_{0}$ in a time scale $\tau$. Along with the lowering of the barrier, atoms begin to leak away from the condensate through the optical lattice. For times $t\gg \tau$, the potential is time-independent and a quasi-stationary regime of leaking can be expected to be reached, at least for sufficiently large condensate reservoirs. Indeed, as discussed later in this section, a quasi-stationary outcoupled black-hole configuration is achieved using some specific protocols. A qualitative scheme of the time evolution of the system is depicted in Fig. \ref{fig:Scheme}.

We describe in the rest of this section the protocol needed to achieve the quasi-stationary regime and its different features, for both {\it ideal} (with a flat envelope) and {\it realistic} (with a Gaussian envelope) optical lattices.

\subsection{Ideal optical lattice}\label{subsec:ideal}

An optical lattice is made of two fixed phase lasers of wavelength $\lambda$ and whose wave vectors form an angle $\theta$ \cite{Fabre2011,Blakie2002}. First, we consider the case of an {\it ideal} finite optical lattice, in which the spatial function $f(x)$ in Eq. (\ref{eq:TDPotential}) reads
\begin{equation}\label{eq:idealOL}
f(x)=\cos^{2}\left[k_L(x-L)\right]\chi\left[\frac{x-\left(L-\frac{d}{2}\right)}{L_{\rm lat}}\right]
\end{equation}
with $k_L=\pi/d$, $d=\lambda/\left[2\sin(\theta/2)\right]$ the lattice period and $\chi(x)$ the characteristic function of the interval $[0,1]$. Specifically, the length of the optical lattice is chosen such that it contains a discrete number of periods, $n_{\rm osc}$, $L_{\rm lat}\equiv n_{\rm osc}d$.

After rescaling the condensate wave function as $\Psi(x,t)\rightarrow\sqrt{n_0}\Psi(x,t)$, the actual degrees of freedom of the system are neatly revealed; once the initial chemical potential is fixed, only $L/\xi_0,\, d/\xi_0,\,\tau/t_{0},\, V_{0}/\mu_0,\, V_{\infty}/\mu_0$
and $n_{\rm osc}$ are free parameters of the problem. However, for typical values of the magnitudes, the features of the system depend very weakly on $(L/\xi_0,\tau/t_{0},V_{0}/\mu_0,n_{\rm osc})$ \cite{deNova2014a}. The major role played by $V_{\infty}/\mu_0$ and $d/\xi_0$ comes from their determining the asymptotic band structure of the optical lattice in the quasi-stationary regime because, as well known, the spectrum of any periodic potential is understood in terms of gaps and energy bands. Throughout this work, when speaking of bands, we will be generally referring to Schr\"odinger (non-interacting) bands, unless otherwise stated. Indeed, as we are dealing with a sinusoidal potential inside the lattice region, the Schr\"odinger equation can be rewritten as a Mathieu's equation, whose properties have been widely studied \cite{Abramowitz1988}. In this simple case, the dimensionless parameter
\begin{equation} \label{small-v}
\zeta\equiv \frac{mV_{\infty}}{8\hbar^2 k_L^{2}}=\frac{V_{\infty}}{16E_R}\propto V_{\infty}d^{2}
\end{equation}
characterizes the asymptotic band structure, $E_R\equiv\hbar^2k^2_L/2m$ being the recoil energy of the optical lattice. For $\zeta\ll1$ we are in the nearly-free atom regime, where bands are wide and gaps are narrow, while for $\zeta\gg1$ we have the opposite tight-binding regime, with narrow and widely spaced bands. In practice, only the lowest energy band is needed to understand the dynamics of the system \cite{deNova2014a}.



In order to provide a quantitative description of the quasi-stationarity of the system, we introduce some useful magnitudes. We define a local chemical potential as
\begin{equation}\label{eq:LocalChemicalPotential}
\mu(x,t)\equiv -\frac{\hbar^{2}}{2m}\frac{\partial_x^{2}\Psi(x,t)}{\Psi(x,t)}+V(x,t)+g|\Psi(x,t)|^{2} \, ,
\end{equation}
For a stationary solution, $\mu(x,t)=\mu$ is real and constant. The one-dimensional current $J(x,t)$ [obtained from Eq. (\ref{eq:GPnormconservation})] is also constant for a 1D stationary solution, as dictated by the continuity equation. However, in the quasi-stationary regime the uniformity of $J(x,t)$ is impossible to fulfill strictly since the current is zero
at $x=0$ due to the hard-wall boundary condition while the leaked flux in the supersonic region obviously carries a non-zero current. Thus, there must be a current gradient and, by the continuity equation, the density has to be time dependent. However, in the
quasi-stationary regime, the condensate leak is sufficiently slow for this time dependence to be weak. It is also reasonable to expect that, once in the quasi-stationary regime, $\mu(x,t)$ is a sufficiently uniform function, with small spatial fluctuations around its average value. In order to measure the homogeneity of $\mu(x,t)$, we define a space-averaged
chemical potential $\bar{\mu}(t)$ and its relative fluctuations $\sigma(t)$ as
\begin{eqnarray}
\bar{\mu}(t) & \equiv  & \frac{\int_{0}^{L_{g}}\mathrm{d}x\,n(x,t)\mu(x,t)}{\int_{0}^{L_{g}}\mathrm{d}x\,n(x,t)}\nonumber \\
\sigma(t) & \equiv  & \frac{1}{\bar{\mu}(t)}
\left[
\frac {\int_{0}^{L_{g}}\mathrm{d}x\,n(x,t)|\mu(x,t)-\bar{\mu}
(t)|^{2}} {\int_{0}^{L_{g}}\mathrm{d}x\,n(x,t)}
\right]^{\frac{1}{2}} \, .
\label{eq:AverageChemicalPotential}
\end{eqnarray}
with $L_g$ the total length of the numerical grid considered. Remarkably, $\mu(x,t)$ and $\bar{\mu}(t)$ can be complex: a non-zero imaginary part of $\mu(x,t)$ reflects a leaking condensate, as revealed by the local relation
\begin{equation}
\partial_t n(x,t)=\frac{2}{\hbar} n(x,t) \, \text{Im} \, \mu(x,t) \, .
\end{equation}
In order to work with strictly real magnitudes, one could also define a local chemical potential by taking the real part of the r.h.s. of Eq. (\ref{eq:LocalChemicalPotential}); that would amount to define $\mu(x,t)$ as the sum of the contributions of the kinetic and potential energies of the flow, and those associated with the local and quantum pressures [see Eqs. (\ref{eq:PhaseAmplitude}) and (\ref{eq:Hydrodynamics})]. Nevertheless, for the characterization of the quasi-stationary regime, both definitions should give similar results due to the smallness of the imaginary part of the chemical potential (see discussion below).

\begin{figure*}[tb!]
\begin{tabular}{@{}cc@{}}
    \includegraphics[width=0.5\columnwidth,valign=b]{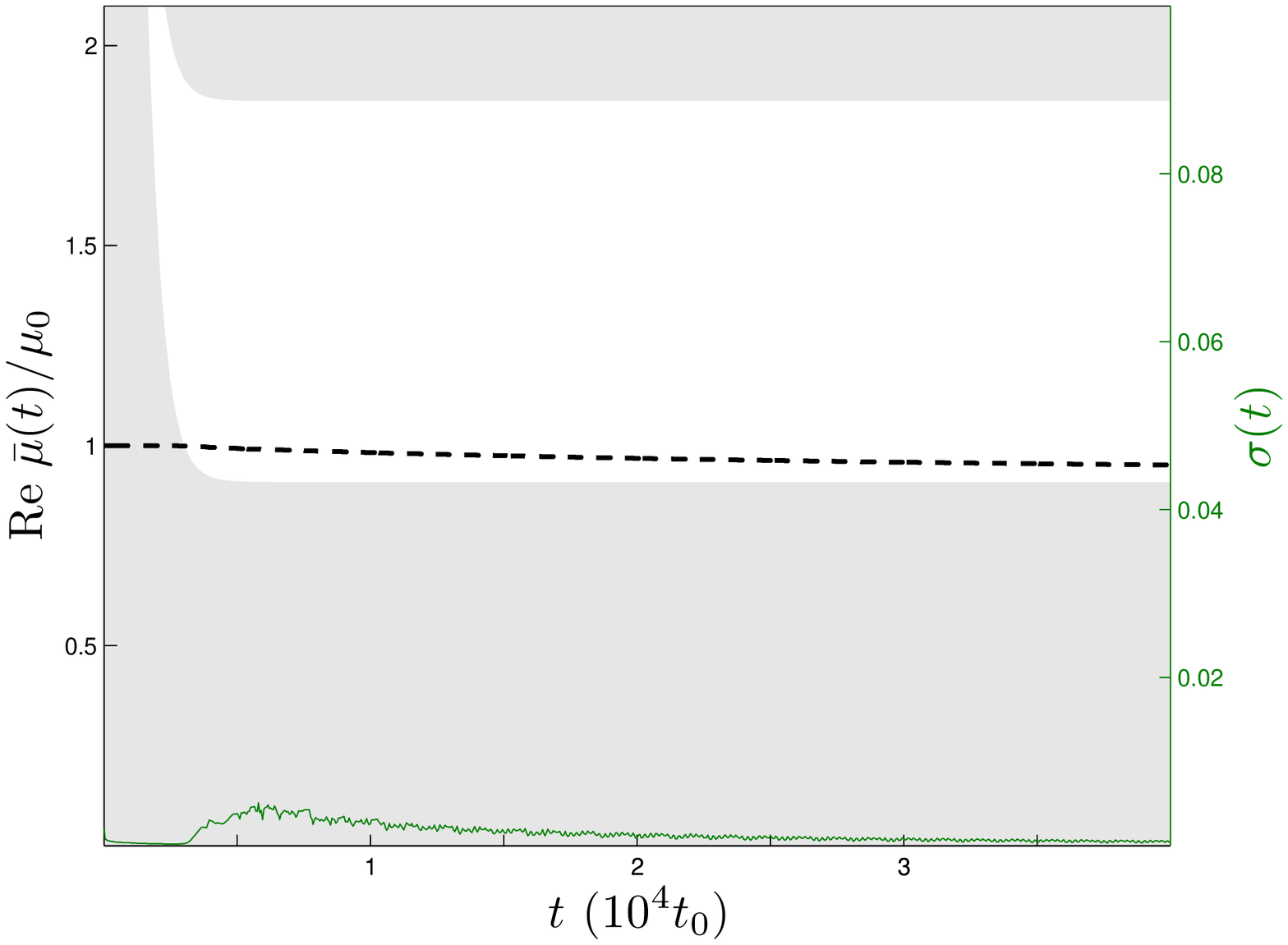} &
    \includegraphics[width=0.49\columnwidth,valign=b]{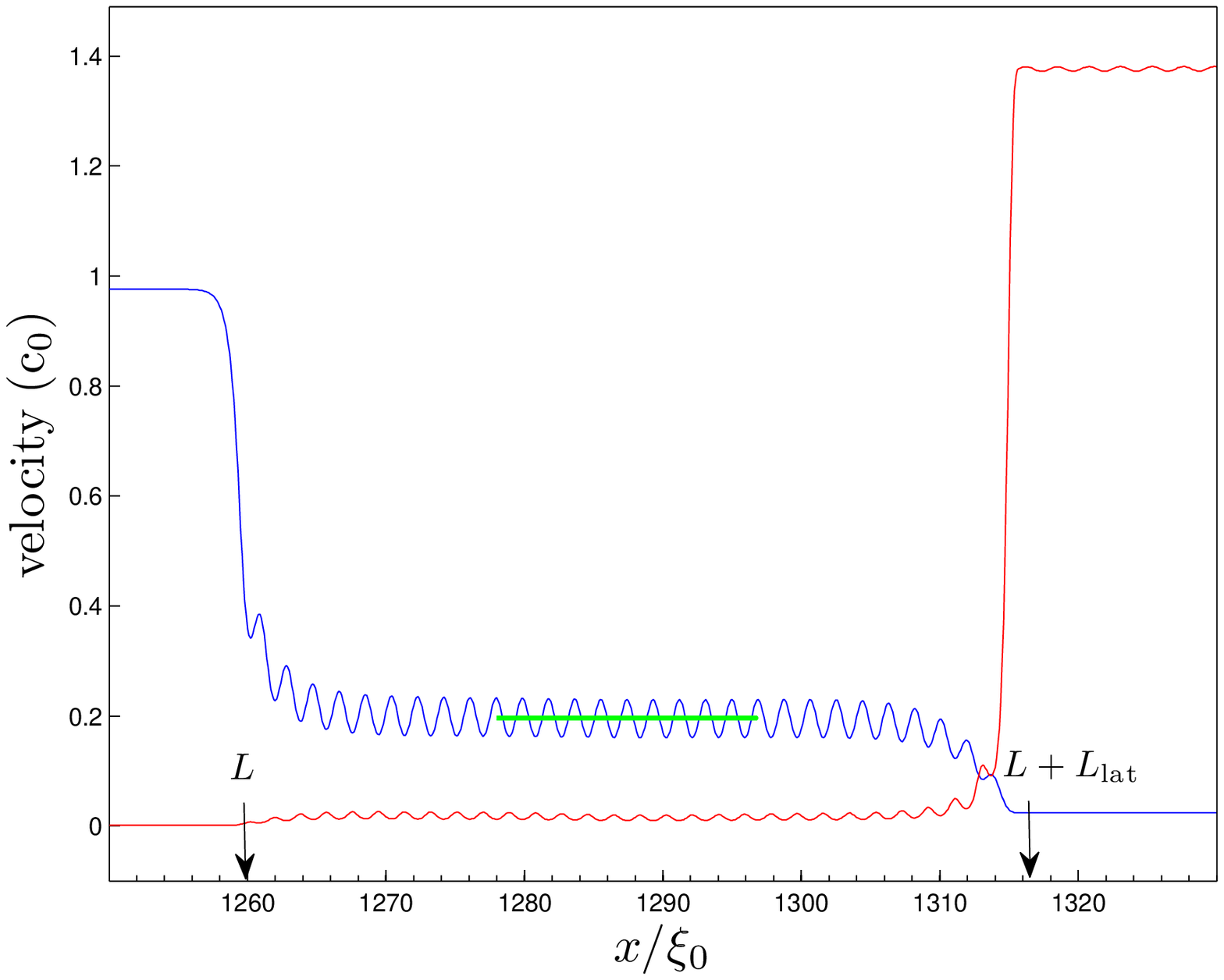} \\
\end{tabular}
\caption{Left plot: Time evolution of the real part of the space-averaged chemical potential $\bar{\mu}(t)$
(black dashed) and its relative fluctuations $\sigma(t)$
(green solid) in an ideal optical lattice. The gap and conduction band of the instantaneous band structure
are indicated, respectively, by grey and white backgrounds.
The values of the parameters are $V_{\infty}=2\mu_0$, $d=1.89\xi_0$, $\tau=500\, t_{0}$,
$L=400\,\mu\text{m}$, $n_{\rm osc}=30$, $N=10^4$, and $\omega_{\rm tr}=2\pi\times4\,\text{kHz}$,
with $n_{0}=25\,\mu\text{m}^{-1}$, $t_0=1.3792\times10^{-4}~\text{s}$ and $\xi_0=0.3175\, \mu\text{m}$.
The value of the dimensionless parameter $\zeta$ that characterizes the asymptotic band structure at large times is $\zeta=0.0905$. Right plot: Local flow velocity (red) and local speed of sound (blue) at a late time
$t=4\times 10^4\, t_{0}$ for the setup of the left plot. The horizontal green segment shows the speed of sound in the optical lattice, Eq. (\ref{eq:latticesound}). Figures taken from Ref. \cite{deNova2014a}.}
\label{fig:IdealBand}
\end{figure*}

The time evolution of $\bar{\mu}(t)$ (black-dashed) and $\sigma(t)$ (solid green) for a typical configuration is shown in left Fig. \ref{fig:IdealBand}, along with the instantaneous band structure computed for the corresponding amplitude $V(t)$ of the potential.
After a transient of order $t\sim 5000 t_0$, all the characteristic magnitudes of the system vary slowly enough in time to consider the system as quasi-stationary. In particular, the real part of the average chemical potential ${\rm Re}\,\bar{\mu}(t)$ drops until it almost reaches the bottom of the asymptotic conduction band, $E_{\rm min}$, ${\rm Re}\,\bar{\mu} \simeq E_{\rm min}$, while its imaginary part, reflecting the leaking of the condensate, is almost negligible, ${\rm Im}\,\bar{\mu}(t)/\mu_0\sim 10^{-6}-10^{-7}$. In fact, the leak is so slow that other limiting processes (for instance, decay of the condensate due to inelastic collisions) present a shorter characteristic time scale. Also, the relative fluctuations of the average chemical potential are of order $10^{-4}$, which can be regarded as a sufficiently small value to consider the system as quasi-stationary.

The required trends for achieving the most favorable quasi-stationary regime as possible can be observed in left Fig. \ref{fig:IdealBand}: (i) initial chemical potential $\mu_0$ slightly above the bottom of the final conduction band, $E_{\rm min}$; (ii) sufficiently wide asymptotic conduction band. This last condition can be understood as the optical lattice acting like a low-pass filter, with the band width being the equivalent of the cutoff frequency, since the higher the cutoff, the wider is the transmission band and then more small waves on top of the condensate travel away through the optical lattice, reducing the spatial fluctuations of the local chemical potential \cite{deNova2014a}.

As $V_{\infty}\gtrsim \mu_0$, wide bands imply that we are in the nearly-free atom limit, $\zeta\ll 1$, in which one can write analytical formulas for the bottom and the top of the lowest energy band, respectively:
\begin{eqnarray}\label{eq:top-bottom}
\nonumber E_{\rm min}(\zeta) & =  & 8 E_R \left[\zeta-\zeta^2+O(\zeta^4)\right]\simeq \frac{V_{\infty}}{2}\\
E_{\rm max}(\zeta)& =  & E_R \left[1+4\zeta-2\zeta^2+O(\zeta^4)\right]\simeq E_R +\frac{V_{\infty}}{4}\, ,
\end{eqnarray}
The width of the band, $\Delta_c\equiv E_{\rm max}-E_{\rm min}$, is then:
\begin{equation}\label{eq:bandwidth}
\Delta_c(\zeta)=E_{\rm max}(\zeta)-E_{\rm min}(\zeta)=E_R \left[1-4\zeta+8\zeta^2+O(\zeta^4)\right]\simeq E_R-\frac{V_{\infty}}{4}
\end{equation}

With respect to the properties of the quasi-stationary flow, right Fig. \ref{fig:IdealBand} shows the velocity profile at a large time $t=4\times 10^4\, t_{0}$. The density profile in the subsonic region is essentially flat in the bulk; this behavior is only modified near the hard-wall at $x=0$, in a scale of a few healing lengths. Indeed, the wave function in the upstream subsonic region is approximated quite well (apart from some trivial phase) by the usual stationary GP solution
\begin{equation}\label{eq:quasistationarywavefunctionsubsonic}
\Psi(x)\simeq \sqrt{n_u}\tanh\left(\frac{x}{\xi_u}\right),~ \xi_u\equiv \sqrt{\frac{\hbar^{2}}{mgn_{u}}}
\end{equation}
with very small fluctuations of both sound and flow velocities on top of it, which implies that $gn_u\simeq \bar{\mu} \simeq E_{\rm min}$ due to the quasi-stationarity of the system, $n_u$ being the subsonic density.

In the optical lattice, the wave function in the bulk corresponds to a Bloch wave and, because of the small value of the density, it is very close to the linear Schr\"odinger limit, which justifies the use of Schr\"odinger bands for the characterization of the system. In particular, the instantaneous value of $\bar{\mu}(t)$ agrees quite well with that computed for an ideal {\it infinite} optical lattice using the average density in the lattice bulk, $\bar{n}_r$, satisfying $\bar{\mu}(t)\simeq E_{\rm min}+O(g\bar{n}_r)\simeq E_{\rm min}$ for a sufficiently small density \cite{deNova2014a}. Nevertheless, within the bulk of the optical lattice the flow is clearly subsonic as can be expected from energetic considerations \cite{deNova2014a}, and the sonic horizon is then placed at its right edge. The sound speed in the optical lattice, $\bar{s}$, is computed in a different way to that of the upstream and downstream regions, although, in the nearly-free atom regime, it can be shown that it is obtained as arising from the average lattice density \cite{deNova2014a}:
\begin{equation}\label{eq:latticesound}
\bar{s}\simeq\sqrt{\frac{g\bar{n}_r}{m}}~.
\end{equation}

Finally, in the downstream region, both density and flow speed profiles are almost uniform and hence they can be characterized by their average values, $n_d(t)$ and $v_d(t)$. As the downstream density is much smaller than in the subsonic region, interactions become negligible and consequently the flow is supersonic. For the same reason, due to the conservation of the chemical potential, $v_d(t)$ remains constant as the quasi-stationary chemical potential is fully transformed into kinetic energy, $\bar{\mu}\simeq mv^2_d/2$.

Apart from gravitational analogue purposes, the achievement of such quasi-stationary regime is of general interest in quantum transport scenarios since it provides a quasi-stationary supersonic current with very well controlled value of the velocity of the atoms as $\bar{\mu}\simeq E_{\rm min}$ and then
\begin{equation}
v_d\simeq \sqrt{\frac{2E_{\rm min}}{m}}
\end{equation}

\subsection{Gaussian-shaped optical lattice}
\label{subsec:Gaussian-shaped}

Here we perform the same analysis as before but using a more realistic Gaussian envelope \cite{Fabre2011,Cheiney2013EPL,Cheiney2013PRA} for the optical lattice potential (\ref{eq:TDPotential}) instead of the flat envelope (\ref{eq:idealOL}),
\begin{equation}\label{eq:actualpotential}
f(x)=\cos^{2}\left[k_L(x-L)\right]\exp\left[-2\left(\frac{x-L}{\tilde{w}}\right)^2\right]
\end{equation}
with $\tilde{w}=w/\cos(\theta/2)$, $w$ being the laser beam width. The magnitude $\tilde{w}$ plays the role of an effective lattice length as $L_{\rm lat}$ for the ideal optical lattice.

\begin{figure*}[tb!]
\begin{tabular}{@{}cc@{}}
    \includegraphics[width=0.5\columnwidth,valign=b]{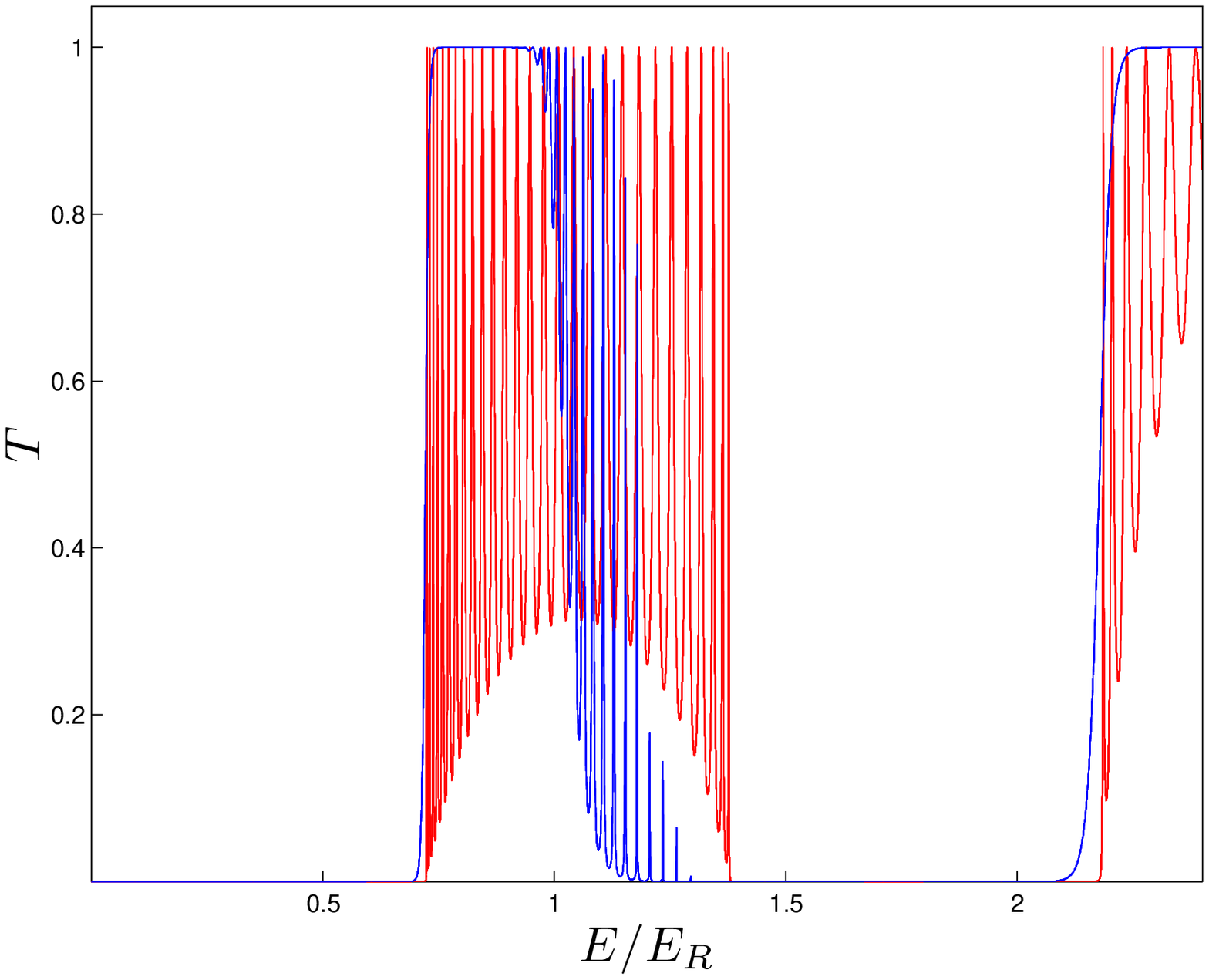} &
    \includegraphics[width=0.51\columnwidth,valign=b]{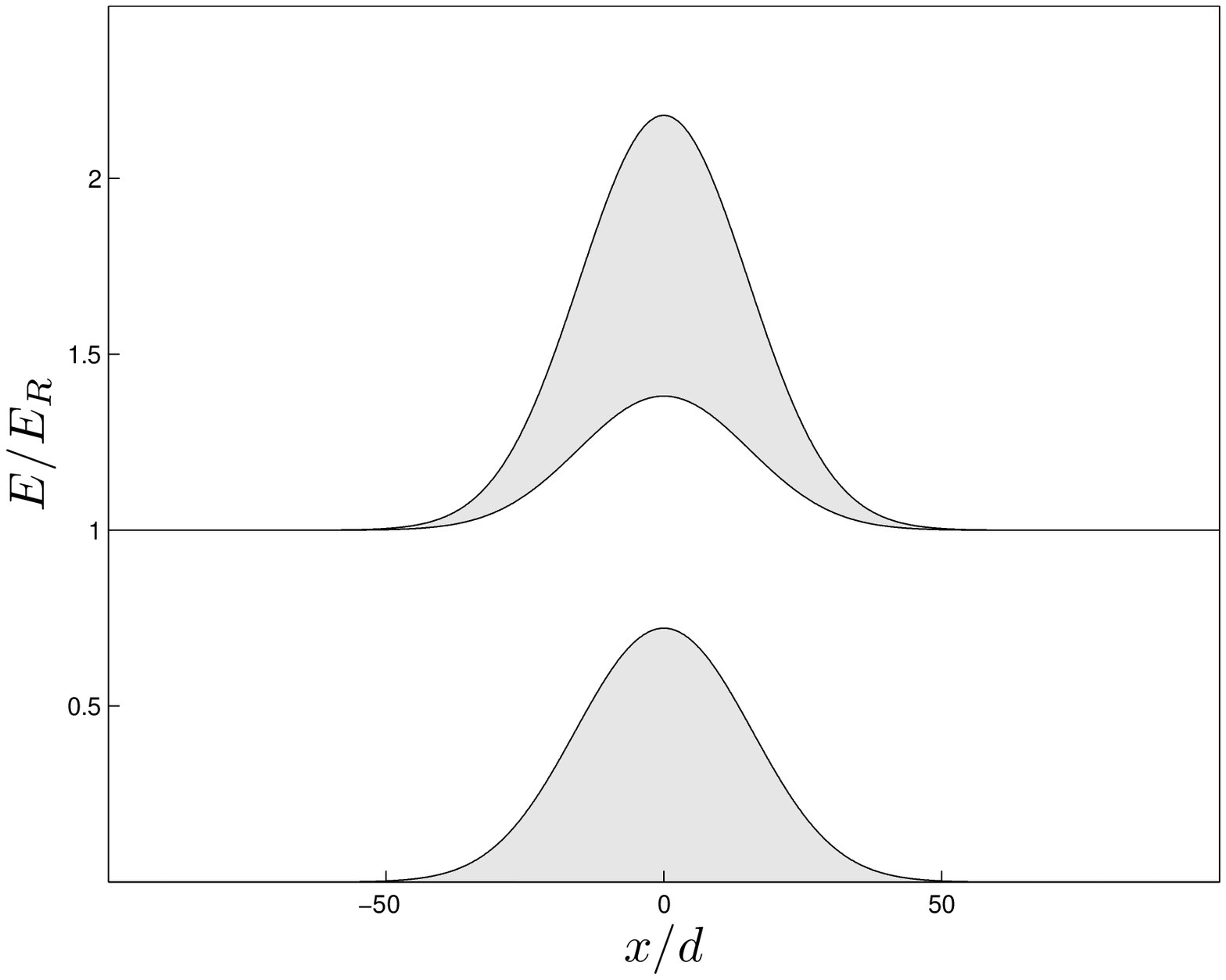} \\
\end{tabular}
\caption{Left plot: Single atom transmission probability $T(E)$, as a function of energy, for a realistic (Gaussian-shaped) optical lattice
(blue) with the instantaneous amplitude $V(t)=1.6 E_R$ and $\tilde{w}=30 d$ and
for an ideal (flat) optical lattice (red) with same amplitude $V(t)$ and $n_{\rm osc}=30$ [see Eqs. (\ref{eq:TDPotential}),(\ref{eq:idealOL}) and (\ref{eq:actualpotential})]. Right plot: Spatially dependent energy bands for the same realistic optical lattice of the left plot. We use the same color criterion for the bands as in Fig. \ref{fig:IdealBand}. Figures taken from Ref. \cite{deNova2014a}.}
\label{fig:RealisticBands}
\end{figure*}

In order to have a sufficiently large condensate reservoir, we require $L\gg \tilde{w}$. On the other hand, for typical laser widths, the Gaussian envelope can be seen as ``adiabatic'', $\tilde{w}\gg d$, which means that the potential $V(x,t)$ can be regarded locally as an {\it ideal} optical lattice at each point of the space with local and instantaneous amplitude $V_{A}(x,t)$ \cite{Santos1998a,Santos1999,Carusotto2000},
\begin{equation}
V(x,t)=V_{A}(x,t)\cos^{2}\left[k_L(x-L)\right],~V_{A}(x,t)\equiv V(t)\exp\left[-2\left(\frac{x-L}{\tilde{w}}\right)^2\right]~,
\end{equation}
Hence, we can expect that the solutions of the linear Schr\"odinger equation associated to this potential show similar properties to those of an ideal optical lattice with the same instantaneous amplitude $V(t)$, as it is confirmed by checking left Fig. \ref{fig:RealisticBands}, where the Schr\"odinger transmission probability is compared for both potentials. In fact, the properties of the Gaussian optical lattice (\ref{eq:actualpotential}) can be understood by locally applying Bloch's theorem. The resulting local band structure is depicted in right Fig. \ref{fig:RealisticBands}: since the bottom of the lowest lattice conduction band is an increasing function of the periodic potential amplitude, the bottleneck for transmission across the realistic lattice occurs at the maximum of its Gaussian envelope. This explains the accurate coincidence between the bottom of both conduction bands shown in left Fig. \ref{fig:RealisticBands}, so $E_{\rm min}$ is still given by Eq. (\ref{eq:top-bottom}). However, for $E>E_R$  the particle encounters a gap somewhere along the Gaussian lattice, which explains the decay for $E>E_R$ of the transmission. Hence, we can take $E_{\rm max}\simeq E_R$ for the realistic lattice, and the bandwidth satisfies:
\begin{equation}\label{eq:bandwidthrealistic}
\Delta_c\simeq E_R-\frac{V_{\infty}}{2}
\end{equation}
Interestingly, in contrast to the more usual peaked resonant structure of the ideal configuration, for $E_{\rm min}(\zeta)<E<E_R$ the realistic setup displays a plateau of essentially perfect atom transmission due to the adiabatic variation of the lattice envelope \cite{deNova2014a}.

\begin{figure*}[tb!]
\begin{tabular}{@{}cc@{}}
    \includegraphics[width=0.5\columnwidth,valign=b]{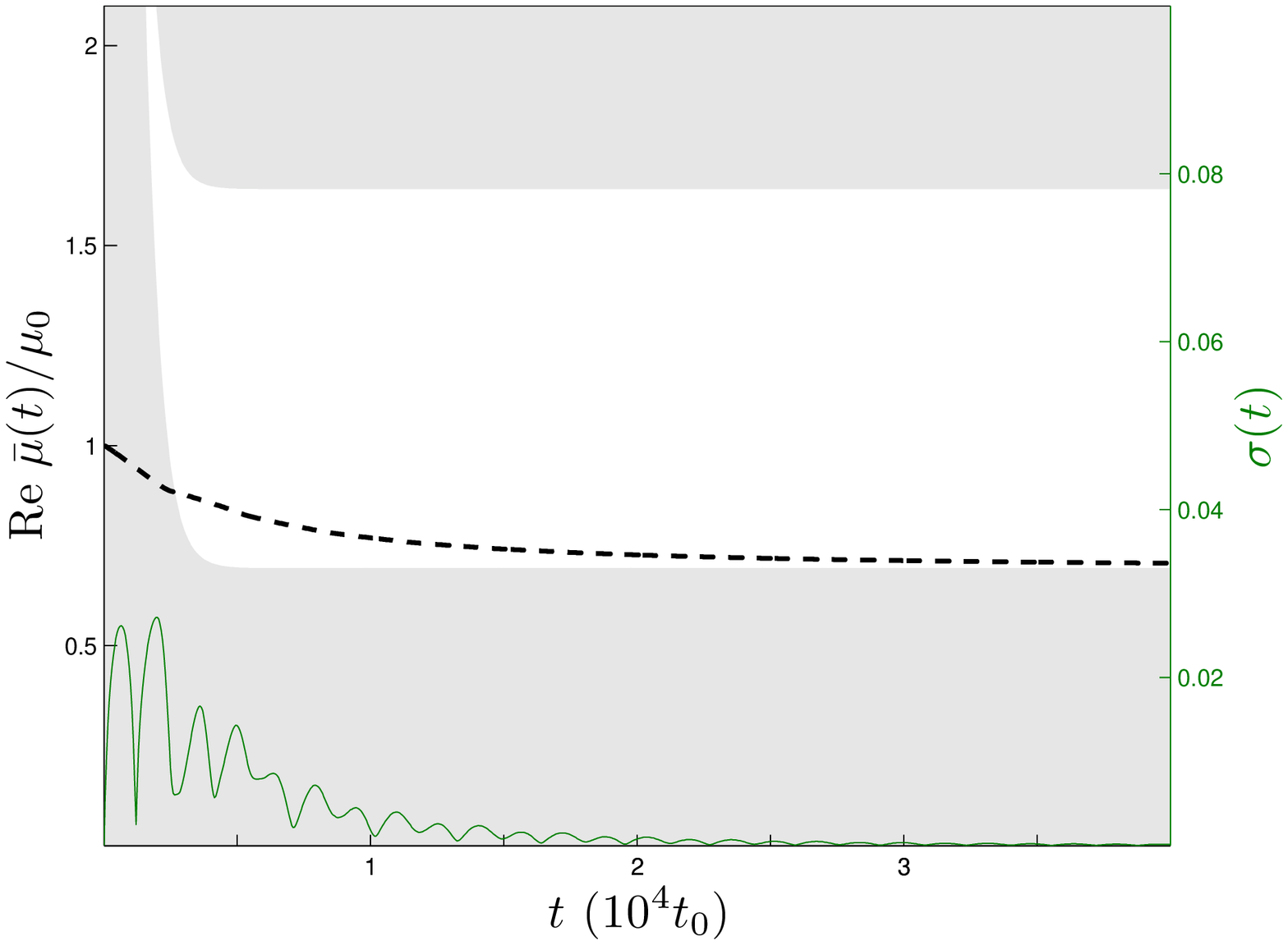} &
    \includegraphics[width=0.49\columnwidth,valign=b]{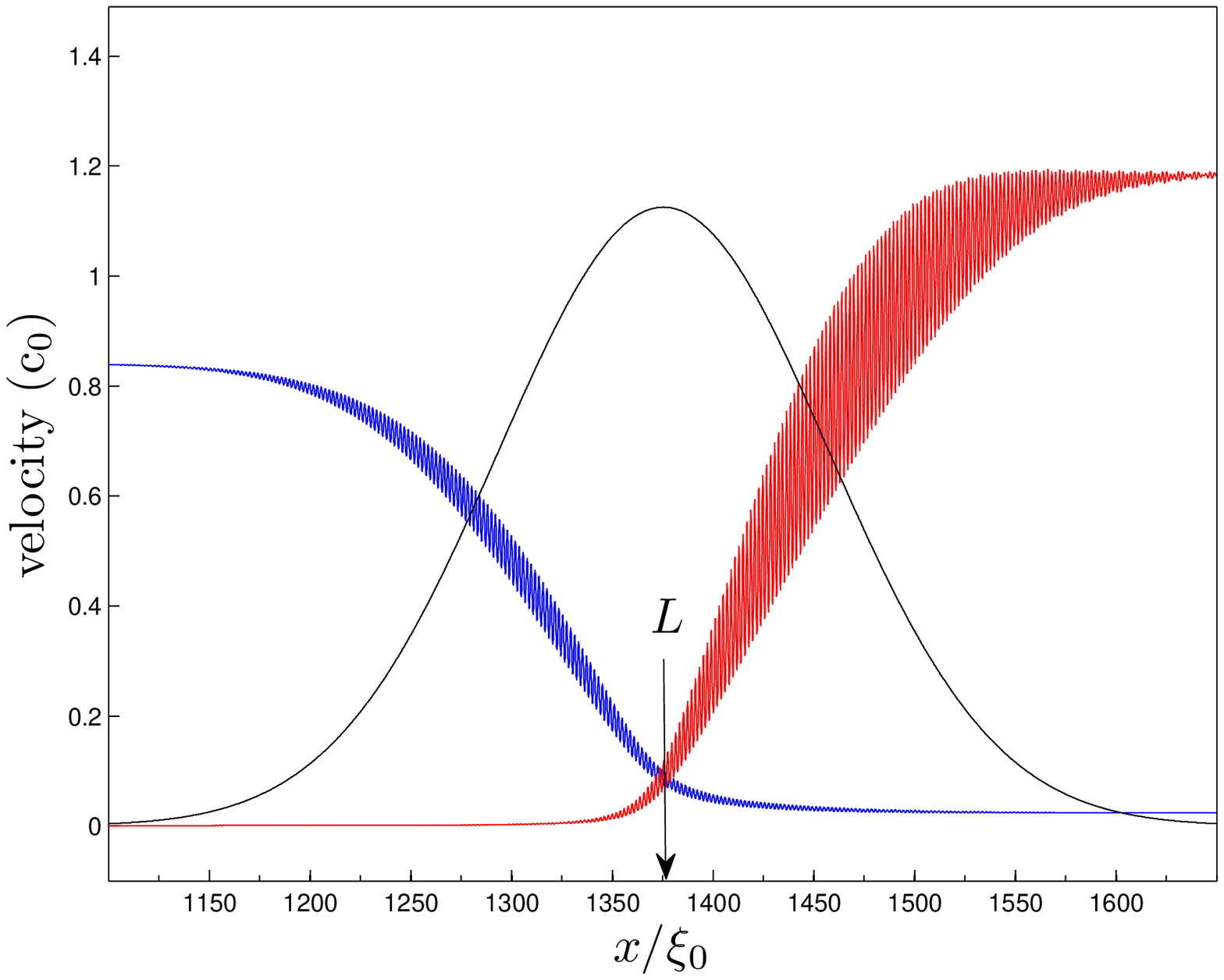} \\
\end{tabular}
\caption{Equivalent of Fig. \ref{fig:IdealBand} but for a realistic optical lattice. Left plot: Time evolution of the real part of the chemical potential and its relative fluctuations. The parameters are  $N=9161$, $L=420~\mu\text{m}$, $\omega_{\rm tr}=2\pi\times 4~\text{kHz}$, $\tilde{w}=50~\mu\text{m}$, $d=600~\text{nm}$, $\tau=500~t_{0}$, $V_{\infty}=1.5~ \mu_0$ and $\xi_0=0.3053~\mu\text{m}$. Right plot: Local flow velocity (red) and local speed of sound (blue) at $t=4\times 10^4\, t_{0}$ for the setup of the left plot. Figures taken from Ref. \cite{deNova2014a}.}
\label{fig:RealisticBand}
\end{figure*}

The requirements of quasi-stationarity are similar to those for an ideal optical lattice: sufficiently broad asymptotic conduction band and initial chemical potential close to its bottom. It is also seen that the quasi-stationary regime is reached when ${\rm Re}~\bar{\mu}(t)$ approaches the bottom of the asymptotic conduction band, that, as discussed above, is the same as that of an ideal optical lattice with equal instantaneous amplitude $V(t)$. All these features are shown in Fig. \ref{fig:RealisticBand}, which is the analog of Fig. \ref{fig:IdealBand} for the realistic case. We see that the reached quasi-stationary regime also presents very small relative fluctuations in the chemical potential, $\sigma(t)\sim 10^{-4}$.

The quasi-stationary profiles of $c(x,t)$ and $v(x,t)$ are shown in right Fig. \ref{fig:RealisticBand}. Once more, in the subsonic and supersonic regions, the density and velocity profiles are essentially flat, as for the ideal optical lattice. In the optical lattice region, the properties of the flow correspond to those of a local oscillating Bloch wave \cite{deNova2014a}. Hence, the general properties of the quasi-stationary flow are similar to those of the ideal case. However, a quite interesting result arises for this realistic configuration. As can be seen in Fig. \ref{fig:RealisticBand}, the sonic horizon is placed right at the maximum of the Gaussian envelope. Of course, this is not a coincidence and it can be shown that indeed it is a general feature of sufficiently smooth potential envelopes \cite{Giovanazzi2004,deNova2014a}.

The birth of the quasi-stationary black hole configuration of Fig. \ref{fig:RealisticBand} can be observed in the following \href{https://www.youtube.com/watch?v=b9YA-Efd4F8}{Movie}, which shows the time evolution of the coarse-grained velocities $\bar{c}(x,t),\bar{v}(x,t)$, that are local averages of the actual sound and flow velocities $c(x,t),v(x,t)$.

Finally, it is worth noting that the described quasi-stationary regime for this realistic configuration can be achieved using typical experimental values for all the parameters of the system \cite{deNova2014a}.

%

\section{Quasi-stationary scattering model}\label{sec:quasistationaryscattering}

We present here the main results of this work: the study of the transport properties on top of the quasi-stationary black-hole regime previously described. Specifically, we compute the scattering $S$-matrix treating the quasi-stationary condensate as a true stationary condensate. From Eqs. (\ref{eq:TDGPOL}), (\ref{eq:LocalChemicalPotential}), it is seen that $i\hbar\partial_t \ln \Psi(x,t)=\mu(x,t)$ and then
\begin{equation}\label{eq:quasistationarywavefunction}
\Psi(x,t)=\Psi(x,t_s)e^{-i\int_{t_s}^{t}\mathrm{d}t'\frac{\mu(x,t')}{\hbar}} \, .
\end{equation}
Therefore, if one chooses $t_s$ when the system is already quasi-stationary, taking the snapshot of the condensate at $t=t_s$ as an actual stationary wave function with chemical potential $\bar{\mu}$ should provide a good approximation as $\mu(x,t)\simeq \bar{\mu}\simeq {\rm Re}\,\bar{\mu}$.

Formally, we consider the time evolution of the fluctuations of the field operator $\hat{\Psi}(x,t)$, defined through $\hat{\Psi}(x,t)=\Psi(x,t)+\hat{\varphi}(x,t)$. Once in the quasi-stationary regime, for times $t>t_s$, we define a quasi-stationary wave function $\Psi_{\infty}(x,t)$ through $\Psi(x,t)\equiv\Psi_{\infty}(x,t)e^{-i\frac{{\rm Re}\,\bar{\mu}}{\hbar}(t-t_s)}$. For simplicity, we neglect in the following the small imaginary part of the chemical potential and take $\bar{\mu}= {\rm Re}\,\bar{\mu}$.

After removing the time-dependent phase associated with the mean chemical potential, $\hat{\varphi}(x,t) \rightarrow \hat{\varphi}(x,t)e^{-i\frac{\bar{\mu}}{\hbar}(t-t_s)}$, and keeping only the lowest order terms [in a similar derivation to that leading to Eqs. (\ref{eq:Fieldequation}), (\ref{eq:TDFieldequation})], the equation governing the time evolution of the fluctuations of the field operator in the Heisenberg picture reads
\begin{eqnarray}\label{eq:TDquasistationaryFieldequation}
\nonumber i\hbar\partial_t\hat{\Phi}&=&M(t)\hat{\Phi},\\
M(t)&=&\left[\begin{array}{cc}G(t) & L(t)\\
-L^{*}(t)&-G(t)\end{array}\right],~\hat{\Phi}(x,t)\equiv\left[\begin{array}{c}\hat{\varphi}(x,t)\\ \hat{\varphi}^{\dagger}(x,t)\end{array}\right]\\
\nonumber G(t)&=&-\frac{\hbar^2}{2m}\partial^2_x+V_{\infty}(x)+2g|\Psi_{\infty}(x,t)|^2-\bar{\mu}\\
\nonumber L(t)&=&g\Psi_{\infty}^2(x,t)~,
\end{eqnarray}
$V_{\infty}(x)\equiv V(x,t=\infty)$ being the asymptotically stationary lattice potential. Now, one can perform a Bogoliubov expansion for $\hat{\Phi}(x,t)$ at $t=t_s$ in terms of the instantaneous eigenstates of $M(t_s)$ and compute their time evolution [see Eq. (\ref{eq:TDBdG}) and related discussion]. The point is that the time variation of the quasi-stationary wave function $\Psi_{\infty}(x,t)$ can be regarded as adiabatic for the ``fast'' modes with instantaneous frequency
\begin{equation}\label{eq:adiabatic}
\bar{\mu}\sigma(t)\ll \hbar\omega \lesssim \bar{\mu},
\end{equation}
Indeed, as seen in Fig. \ref{fig:RealisticBand}, the typical time scale of variation of the quasi-stationary regime is $\sim 10^4t_0\sim 1~\text{s}$, much larger than the actual duration of the current experiment of Ref. \cite{Steinhauer2016}, $\sim 0.1\text{s}$. Thus, in order to study the relevant properties of the system in the quasi-stationary regime near a time $t=t_s$, one can approximate $\Psi_{\infty}(x,t)\simeq \Psi_{\infty}(x,t_s)=\Psi(x,t_s)\equiv \Psi_{\infty}(x)$.

\begin{figure*}[tb!]
\includegraphics[width=1\textwidth]{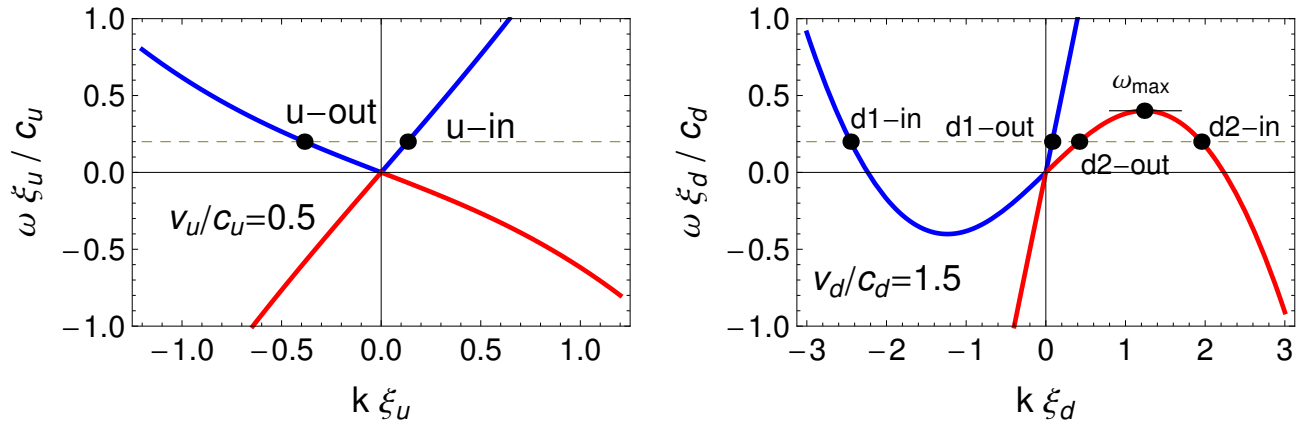}
\caption{Dispersion relation in the asymptotic homogeneous subsonic and supersonic regions, as given by Eq. (\ref{eq:dispersionrelation}). Left: dispersion relation in the subsonic region, denoted by label $u$. Right: dispersion relation in the supersonic region, denoted by label $d$. The blue (red) curves represent positive (negative)
normalization of the modes. Here, $\xi_{u(d)}$ are the
corresponding healing lengths, $c_{u(d)}$ and $v_{u(d)}$ the corresponding sound and
flow velocities, respectively, while $\omega_{{\rm max}}$ is the threshold frequency of the anomalous channel $d2$.}
\label{fig:DispRelation}
\end{figure*}

As a result of the previous considerations, modes satisfying condition (\ref{eq:adiabatic}) are also quasi-stationary and hence given by the solutions of the {\it stationary} Bogoliubov-de Gennes (BdG) equation
\begin{equation}\label{eq:effectiveTIBdG}
\hbar\omega_n z_{n}=M_\infty z_{n},~M_\infty \equiv M(t_s),~
\end{equation}
with $n$ some index labeling the eigenstates of the spectrum. Thus, we only need to compute the quasi-stationary states $z_{n}$ in order to characterize the quantum fluctuations around the mean-field condensate. In our case, as we have two asymptotic homogeneous regions, the subsonic and the supersonic region (see Figs. \ref{fig:IdealBand}, \ref{fig:RealisticBand}), we can regard the calculation as a scattering problem; the corresponding dispersion relation for each region is represented in Fig. \ref{fig:DispRelation}. As well known from scattering theory, the associated spectrum is continuous, formed by scattering states with energy $\hbar\omega$ and labeled as $z_{a,\omega}(x,t_s)$, with $a$ the incoming channel, $a=u$ labeling the subsonic modes and $a=d1,d2$ the normal (anomalous) modes in the supersonic region [check also Appendix \ref{app:generalformalism} for more details]. In particular, all the relevant information is contained in the scattering matrix so we focus on its computation.

For solving the scattering problem, we consider the subsonic and supersonic regions as semi-infinite and approximate the wave function there as a perfect plane wave, $\Psi_{\infty}(x)=\sqrt{n_i}e^{iq_ix}e^{i\theta_i}$, $i=u,d$, which is indeed a very good approximation as explained in Sec. \ref{sec:themodel}. The corresponding sound speed and flow velocity are $c_i=\sqrt{gn_i/m}$ and $v_i=\hbar q_i/m$, respectively. Specifically, from Eq. (\ref{eq:quasistationarywavefunctionsubsonic}), we see that in the plateau of the subsonic region the value of $n_u$ is related to the average chemical potential as $\bar{\mu}=gn_u$, giving a sound velocity $c_u=\sqrt{gn_u/m}$, while the momentum of the plane wave can be taken as zero, $q_u=0$, due to the small value of the flow velocity. In the supersonic region, the density is $n_d$ and the flow velocity is $v_d$, so $c_d=\sqrt{gn_d/m}$ and $q_d=mv_d/\hbar$. Indeed, the supersonic Mach number $v_d/c_d$ is quite large and the corresponding BdG solutions are very close to free atom plane waves. Within this scheme, the $S$-matrix is obtained following the procedure described at the end of Appendix \ref{app:generalformalism}. In the light of the experimental evidence of Ref. \cite{Steinhauer2016}, this model should be valid to study the transport properties of the system.

We remark that the current associated to the quasi-stationary GP wave function $\Psi_{\infty}(x)$ is non-uniform since the condition of uniformity is impossible to fulfill in this setup as explained before. This should not represent a problem as in the actual experimental setup of Ref. \cite{Steinhauer2016} the current was also non-uniform, presenting large differences in its value between both asymptotic regions. Nevertheless, the quasi-particle current associated to Eq. (\ref{eq:effectiveTIBdG}) is indeed homogeneous, which guarantees the pseudo-unitarity of the $S$-matrix  [see Eqs. (\ref{eq:stationarycurrentssolutions}), (\ref{eq:pseudounitarity})].

As expected from Eq. (\ref{eq:adiabatic}), the previous stationary BdG approximation necessarily fails in the limit $\omega\rightarrow 0$ since in that regime the weak time-dependence of the wave function cannot be neglected. In particular, the zero-energy Goldstone mode arising from phase transformations of the GP wave function [see Eq. (\ref{eq:zeromode})] is not anymore a solution of the quasi-stationary BdG equations (\ref{eq:effectiveTIBdG}) as $\Psi_{\infty}(x)$ is not strictly stationary. A possible way to overcome this problem is to make use of the {\it relative} BdG equations \cite{Macher2009a}, instead of the usual BdG equations here considered, where one uses an expansion for the field operator of the form $\hat{\Psi}(x,t)=\Psi(x,t)[1+\hat{\phi}(x,t)]$, $\hat{\phi}(x,t)$ being the relative fluctuations of the field operator. However, due to the non-perfect stationarity of the GP wave function, the corresponding relative quasi-particle current for the quasi-stationary BdG modes is also inhomogeneous, which implies that the corresponding $S$-matrix is not pseudo-unitary. Nevertheless, since the region near $\omega=0$ is not interesting for the observation of the spontaneous Hawking effect \cite{deNova2014,Busch2014,deNova2016}, for practical purposes we can restrict the computations to the range $\omega > \omega_{\Lambda}$, $\omega_{\Lambda}$ being a numerically enforced cut-off of order $\omega_{\Lambda}\sim 10^{-2}\mu_0 \gg \sigma(t)$ so condition (\ref{eq:adiabatic}) is fulfilled and the stationary BdG approximation (\ref{eq:effectiveTIBdG}) is expected to be valid.

In the actual numerical computation of the scattering matrix a problem arises because of solutions whose amplitude grows exponentially along the optical lattice; this behavior can be understood in terms of local Bloch waves with complex wave vector. Due to the relatively large size of the optical lattice, these exploding modes give rise to singular fundamental matrices (within computer's accuracy), spoiling the computation of the scattering matrix. This problem is known as the {\it large fd problem} \cite{Lowe1995} and appears in a wide variety of fields. We discuss in Appendix \ref{app:numerical} two different methods to deal with it, using the Global Matrix method (appearing in the study of the propagation of ultrasonic waves in multilayered media \cite{Lowe1995}) or the QR decomposition (considered in Anderson localization problems \cite{Slevin2004}).

Before the presentation of the numerical results, some general remarks regarding the structure of the spectrum are in order. First, we note that in the quasi-stationary regime the chemical potential is placed near the bottom of the conduction band, so $\bar{\mu}\simeq E_{\rm min}$. Moreover, due to the large supersonic Mach number, the sound speed in the dispersion relation can be neglected and then $\omega_{\rm max}$ (the threshold above which the anomalous channel $d2$ disappears, see Fig. \ref{fig:DispRelation}) satisfies $\hbar\omega_{\rm max}\simeq mv_d^2/2 \simeq \bar{\mu}$. On the other hand, the width of the conduction band of the BdG solutions, $\Delta^{\rm BdG}_c$, satisfies in this limit $\Delta^{\rm BdG}_c \simeq \Delta_{c}$, with small corrections $O(g\bar{n}_r)$ due to the quadratic interacting terms containing the GP wave function in the corresponding BdG equations \cite{deNova2014a}. Thus, one can distinguish in principle two qualitatively different regimes depending on whether the whole BdG conduction band is contained within the frequency range of the Hawking spectrum, $\Delta_c<\hbar\omega_{\rm max}\simeq E_{\rm min}$, which implies
\begin{equation}\label{eq:regimegeneral}
E_{\rm max}\lesssim 2E_{\rm min}
\end{equation}
Using Eqs. (\ref{eq:top-bottom}), (\ref{eq:bandwidth}), this condition reads for the ideal optical lattice as:
\begin{equation}\label{eq:regimeideal}
E_R\lesssim \frac{3}{4}V_{\infty}
\end{equation}
while for the realistic optical lattice it is simply reduced to
\begin{equation}\label{eq:regimegaussian}
E_R\lesssim V_{\infty}
\end{equation}
We remark that the above relations are not exact but rather approximate estimations which indeed describe quite well the qualitative features.

\subsection{Ideal optical lattice}

\begin{figure}[!tb]
\begin{tabular}{@{}cc@{}}
    \includegraphics[width=0.5\columnwidth,valign=b]{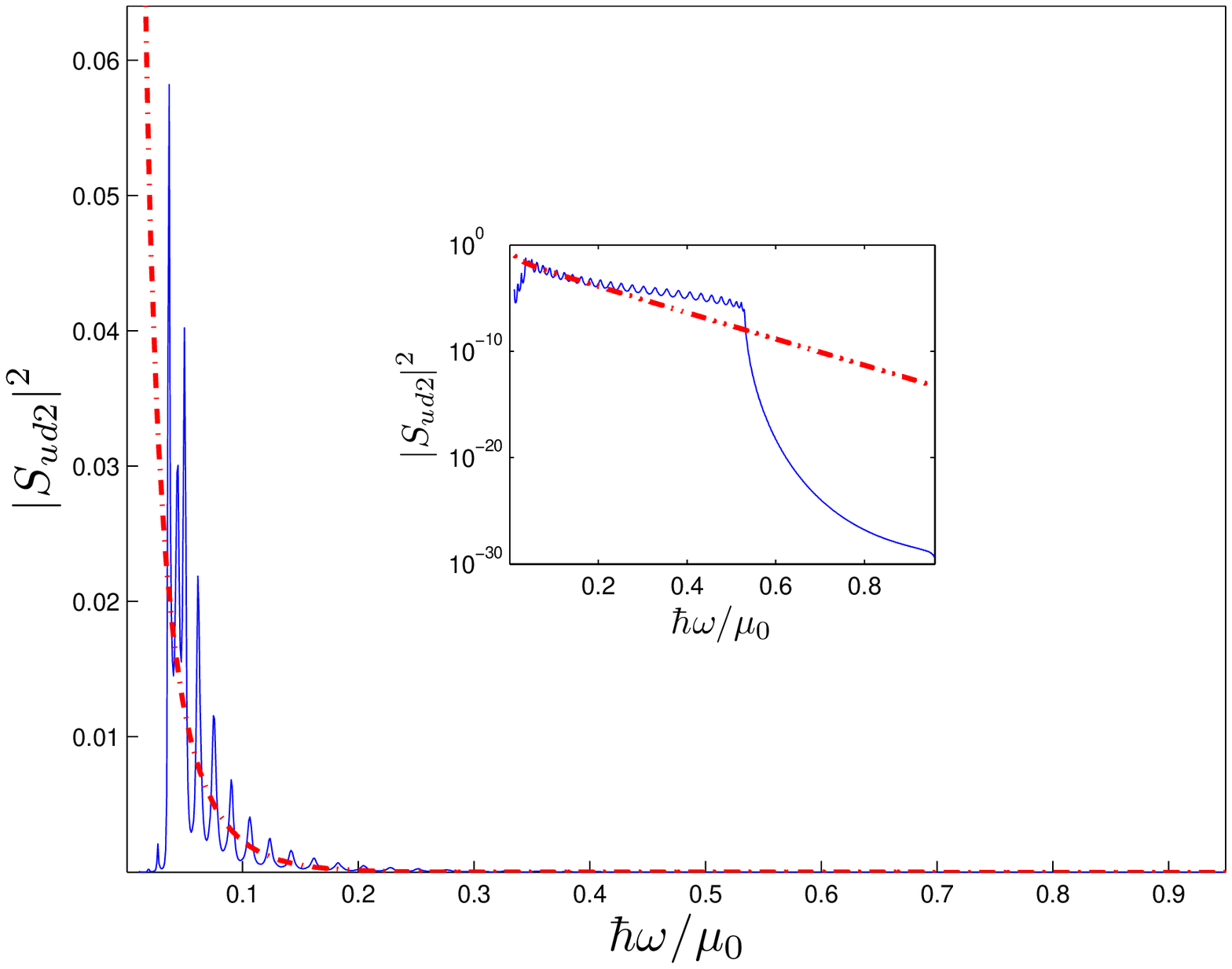} & \includegraphics[width=0.51\columnwidth,valign=b]{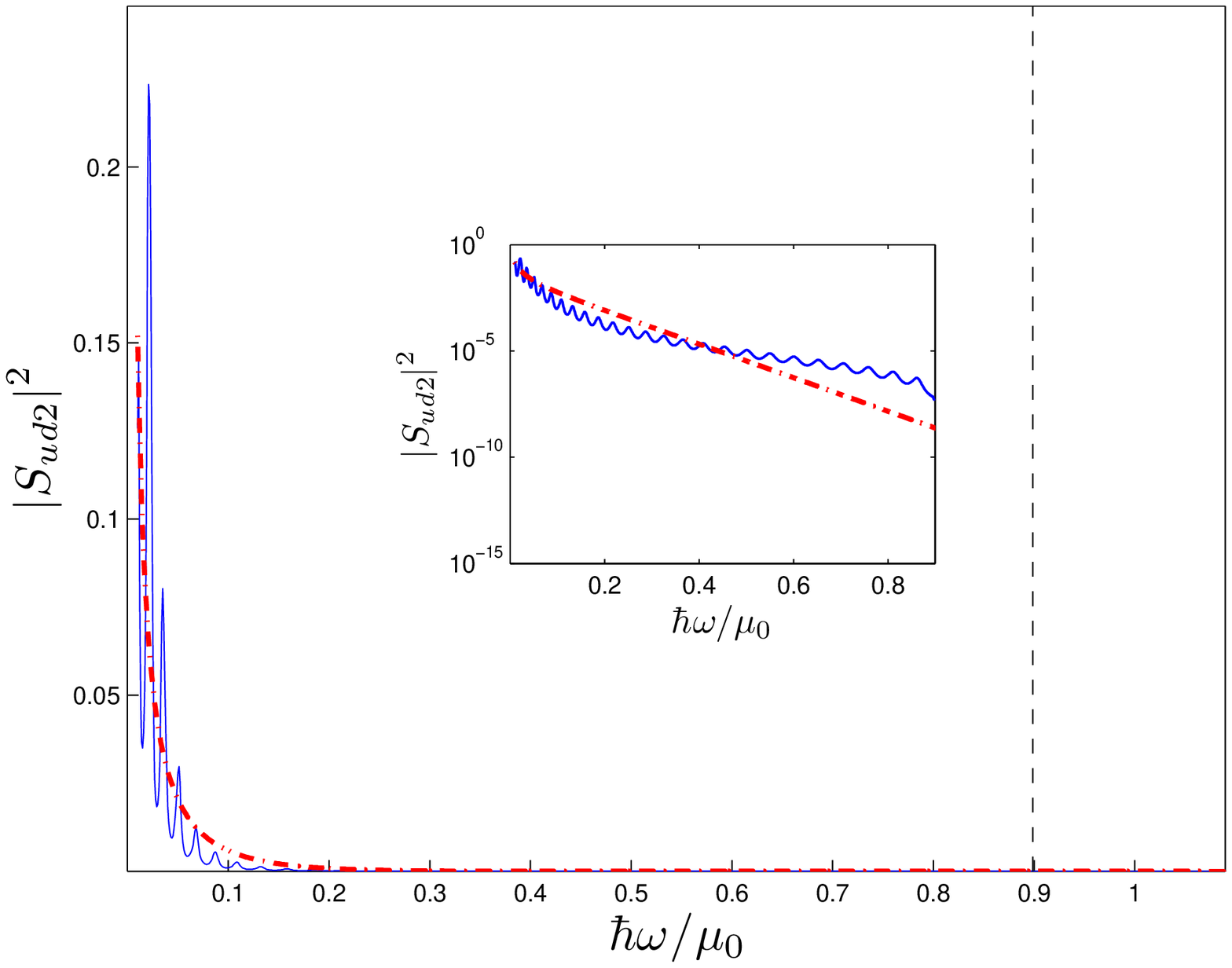} \\
    \includegraphics[width=0.5\columnwidth,valign=b]{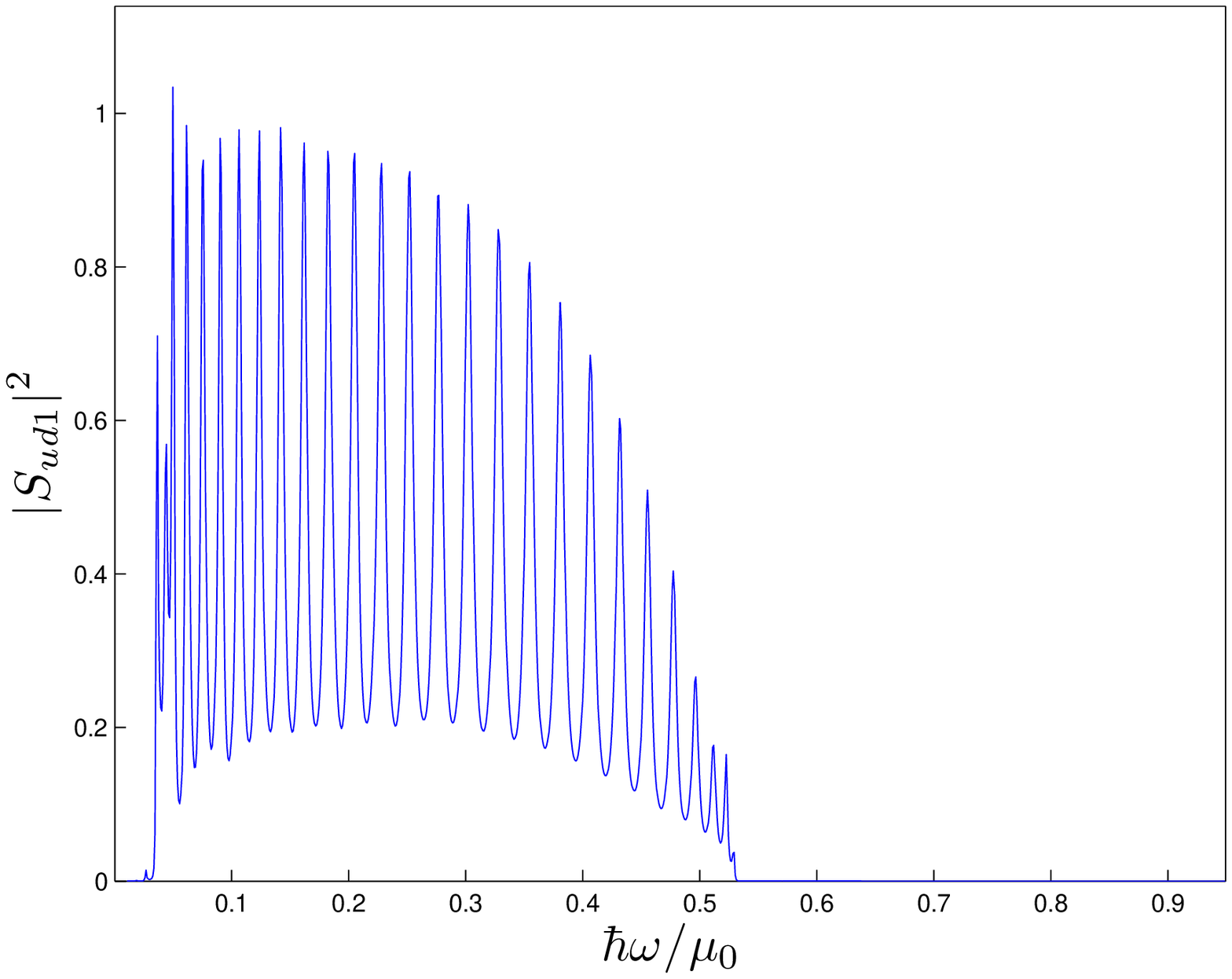} & \includegraphics[width=0.51\columnwidth,valign=b]{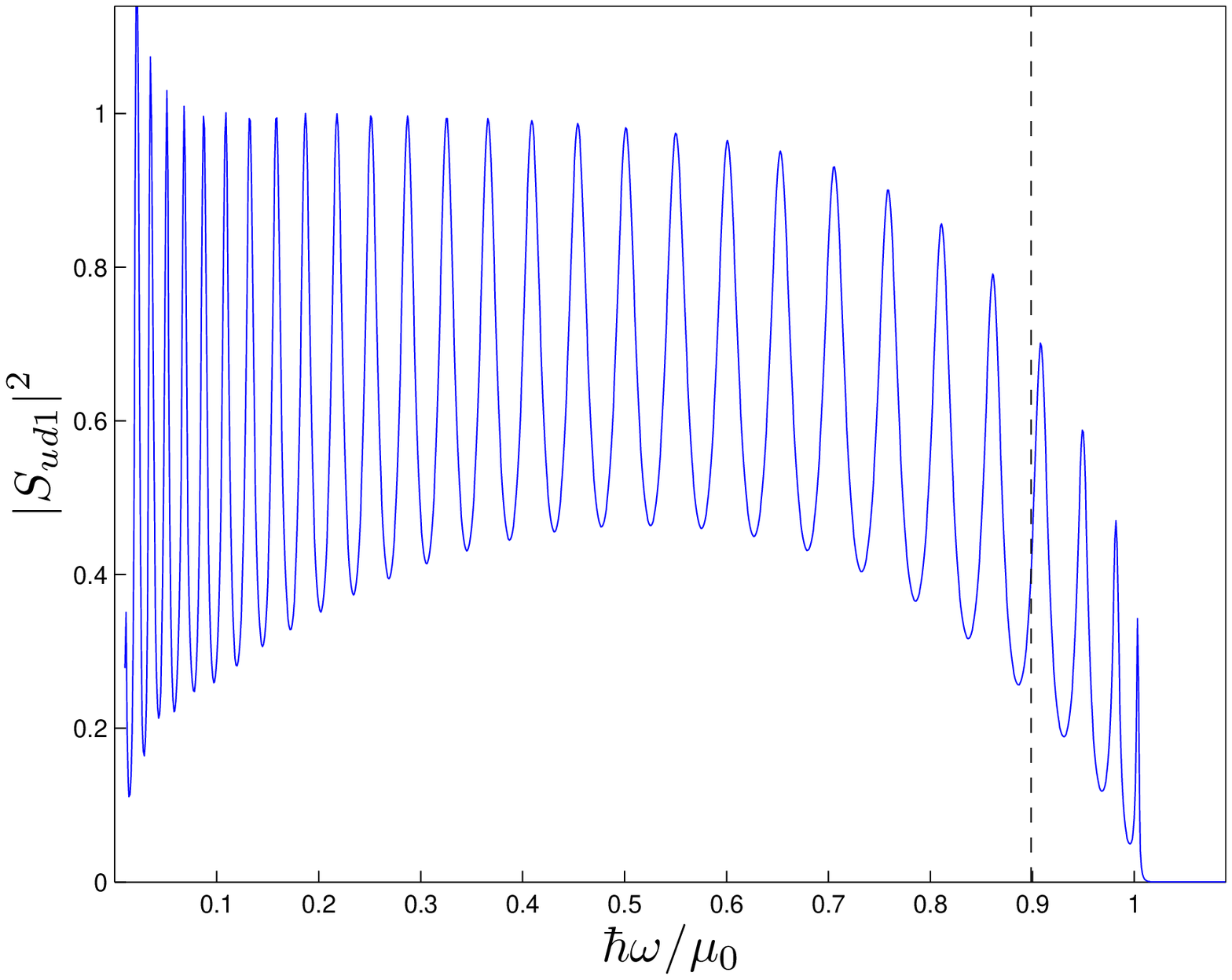}
\end{tabular}
\caption{Hawking spectrum (solid blue line) for two different quasi-stationary black-hole configurations in an ideal optical lattice. The BdG solutions are computed
on top of the quasi-stationary mean-field background provided by the wave function evaluated at a time $4\times 10^4\, t_{0}$.
For both columns, $\tau=500\, t_{0}$,
$L=400\,\mu\text{m}$, $n_{\rm osc}=30$, $N=10^4$, and $\omega_{\rm tr}=2\pi\times4\,\text{kHz}$, and then
$n_{0}=25\,\mu\text{m}^{-1}$, $\xi_0=0.3175\,\mu\text{m}$ and $t_0=1.3792\times10^{-4}~\text{s}$,
only varying the long-time potential amplitude $V_{\infty}$ and the lattice spacing $d$. The dashed-dotted red line represents a fit of the data to a Planck distribution (\ref{eq:Planck}). We note that we have
introduced a numerical infrared cutoff $\hbar\omega_{\Lambda}=0.01\mu_0$ to avoid the low frequency region. The insets in the upper row are the same plots but in logarithmic scale. Left column: $V_{\infty}=2.13~\mu_0$ and $d=2.36\xi_0$, which gives $E_{\rm min}=0.91~\mu_0$, $E_{\rm max}=1.38~\mu_0$ and $\Delta_c=0.47~\mu_0$.
The long time parameters of the quasi-stationary flow are $\bar{\mu}\simeq \omega_{\rm max}=0.96~\mu_0$, $c_u=0.98~c_0$, $v_d=1.39~c_0$ and $c_d=0.02~c_0$.
Right column: $V_{\infty}=1.86~\mu_0$ and $d=1.91\xi_0$, which gives $E_{\rm min}=0.85~\mu_0$, $E_{\rm max}=1.8~\mu_0$ and $\Delta_c=0.95~\mu_0$. The long time parameters of the quasi-stationary flow are $\bar{\mu}\simeq \omega_{\rm max}=0.90~\mu_0$, $c_u=0.95~c_0$, $v_d=1.34~c_0$ and $c_d=0.03~c_0$. The vertical dashed line represents the threshold frequency $\omega_{\rm max}$.}
\label{fig:BdGOLIdeal}
\end{figure}

First, we compute the scattering matrix for the quasi-stationary flow of the ideal optical lattice of Sec. \ref{subsec:ideal}, using as background the quasi-stationary wave function $\Psi_{\infty}(x)=\Psi(x,t_s)$, with $t_s=4\times 10^4\, t_{0}$. In Fig. \ref{fig:BdGOLIdeal}, the two qualitatively different scenarios previously distinguished are shown. In the left column, we show the results corresponding to a situation in which the BdG conduction band is completely contained inside the Hawking radiation frequency range, $\Delta_c<\hbar\omega_{\rm max}$, while the right column is devoted to the opposite regime, $\Delta_{c}>\hbar\omega_{\rm max}$. In the upper row, we show the anomalous-normal transmission, $|S_{ud2}|^2$, which characterizes the spontaneous Hawking emission [see Eq. (\ref{eq:hawkingef}) and related discussion]. In the lower row, we plot the normal-normal transmission element $|S_{ud1}|^2$, corresponding to an incident quasi-particle in the normal scattering channel $d1$ transmitted to the outgoing upstream channel. We have not observed any essential difference in the results when taking the snapshot of the condensate at a different time once the system is quasi-stationary.

The Hawking spectrum $|S_{ud2}|^2$ shows a peaked structure, with a decaying envelope similar to that of non-resonant spectra, which are known to be approximated quite well in a wide range of frequencies by a Planck distribution \cite{Macher2009a,Larre2012,Busch2014,Michel2016a}
\begin{equation}\label{eq:Planck}
f(\omega)=\frac{A}{e^{\frac{\hbar\omega}{k_BT_H}}-1}
\end{equation}
with $A$ some grey-body factor; a fit of the Hawking spectrum to $f(\omega)$ is plotted as a dashed-dotted red line in the upper row of Fig. \ref{fig:BdGOLIdeal}. When comparing the columns in more detail, we observe the behavior anticipated in the discussion leading to Eq. (\ref{eq:regimegeneral}): the left column displays a highly non-thermal behavior as the whole conduction band is contained in the Hawking spectrum, creating a sharp cut-off in the transmission band above which the transmission becomes exponentially small (see in particular the logarithmic plot in the insets of the upper row) in a similar way to the Schr\"odinger transmission of Fig. \ref{fig:RealisticBands}.

By comparing both rows, we clearly see that the transmission band is greatly dominated by normal-normal processes while the anomalous-normal transmission plays a secondary role. This effect can be understood by further inspecting the structure of the BdG solutions in the optical lattice and in the supersonic region. Due to the low value of the interacting term in both regions, for frequencies satisfying  $\hbar\omega \gg g\bar{n}_r, gn_d$, the mixing of the $u,v$ components of the BdG spinors is very small (check Appendix \ref{app:generalformalism} for more details about the notation and structure of the BdG equations). As the optical lattice is subsonic, the modes corresponding to propagating Bloch waves have positive normalization and thus present a dominant $u$ component and an associated small $v$ component. In the supersonic region, the mixing between both components is indeed extremely small due to the large value of the Mach number; while the $d1$ modes have dominant $u$ component, the anomalous $d2$ modes are those with dominant $v$ component. Then, when matching the solutions in the optical lattice with those in the supersonic region, the normal-normal transmission dominates over the anomalous-normal transmission. Note that, due to the presence of an anomalous channel, the normal-normal transmission can be larger than the unity as can be seen from the pseudo-unitary relation $|S_{uu}(\omega)|^2+|S_{ud1}(\omega)|^2-|S_{ud2}(\omega)|^2=1$ [see Eq. (\ref{eq:pseudounitarity}) and related discussion].

\begin{figure}[!tb]
\begin{tabular}{@{}cc@{}}
    \includegraphics[width=0.5\columnwidth,valign=b]{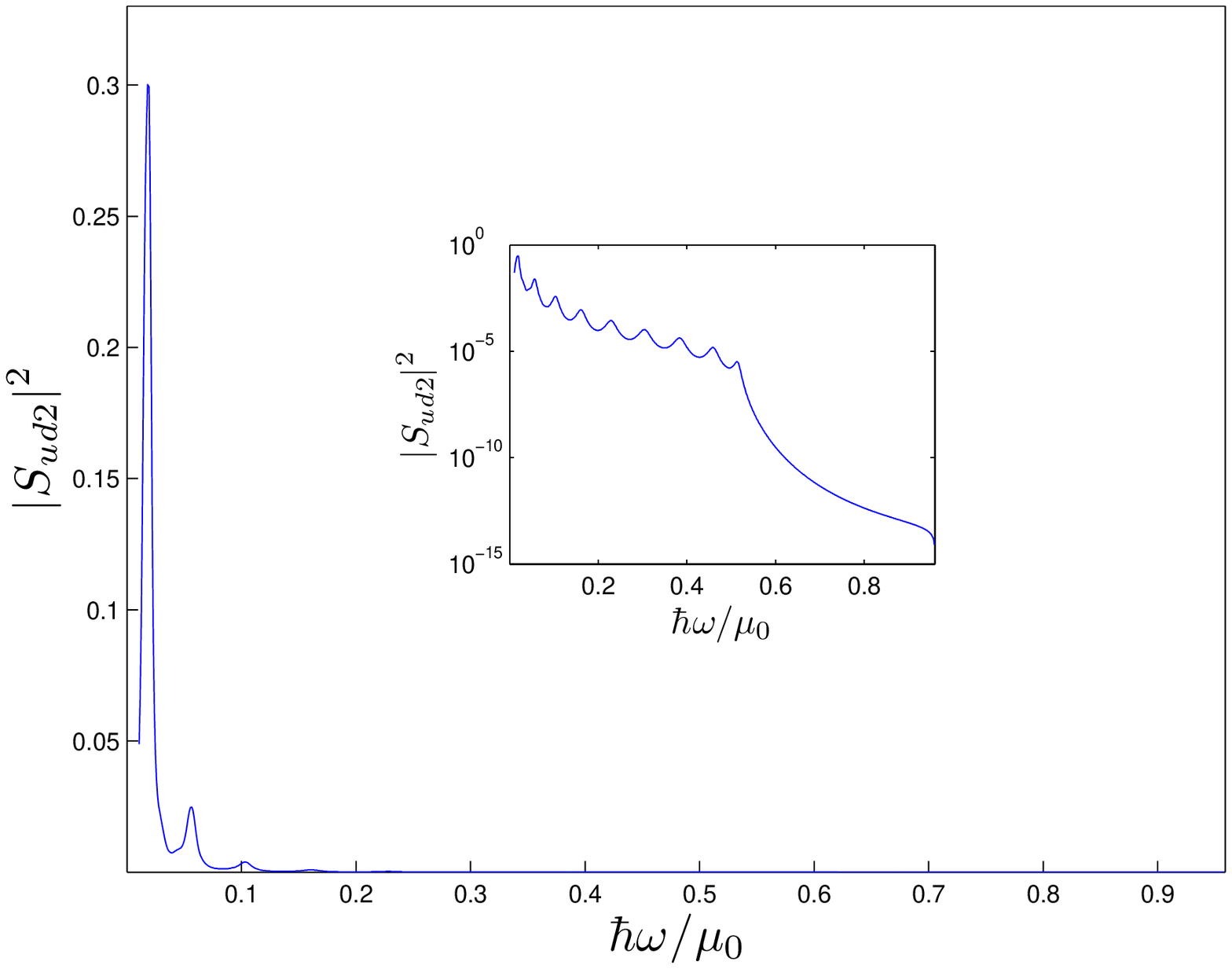} & \includegraphics[width=0.51\columnwidth,valign=b]{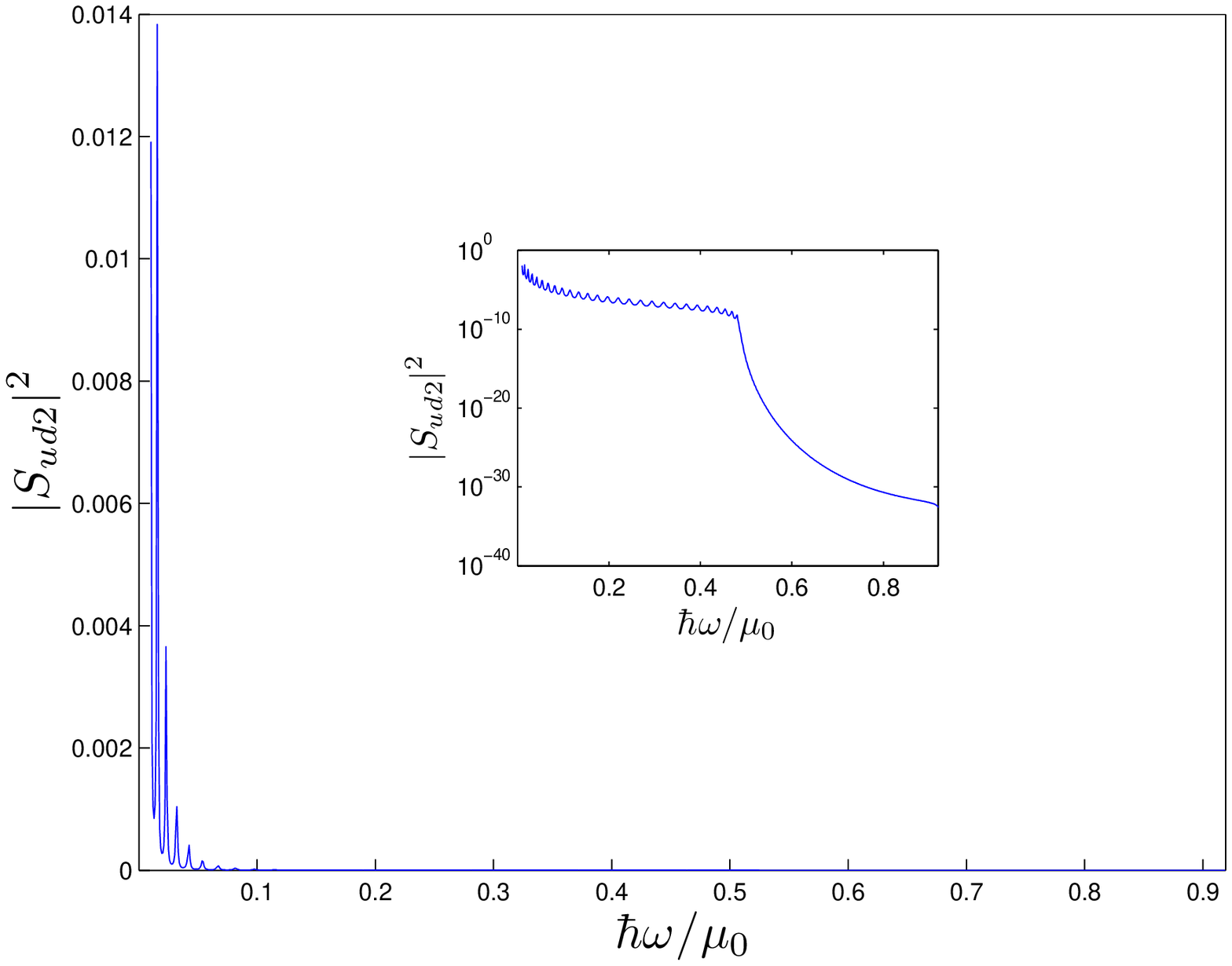}
\end{tabular}
\caption{Same as Fig. \ref{fig:BdGOLIdeal} for two configurations with the same parameters as that of its left column, but now for a lower number of oscillations $n_{\rm osc}=10$ (left figure) and for a smaller condensate ($N=1000$ and $L=40\,\mu\text{m}$). We note that the value of the supersonic speed of sound in the latter case is as low as $c_d=0.006~c_0$, almost an order of magnitude smaller than in Fig. \ref{fig:BdGOLIdeal}.}
\label{fig:BdGOLIdealComparisons}
\end{figure}

We study in Fig. \ref{fig:BdGOLIdealComparisons} the influence of other parameters on the Hawking spectrum. In the left plot, we show a simulation with the same parameters as those of the left column of Fig. \ref{fig:BdGOLIdeal} but with a reduced number of oscillations in the lattice, $n_{\rm osc}=10$, which results in a less peaked spectrum and a lower exponential suppression of the transmission, as expected from the usual results of conventional scattering theory. In the right plot, instead of reducing the number of peaks in the lattice, we consider a smaller confined condensate, observing a decrease of the intensity of the Hawking spectrum. This feature can be understood as follows: a smaller reservoir provides a lower value of the escaping current and thus, the value of the density is reduced in the supersonic region. By virtue of the arguments given above, a smaller interacting term in the supersonic region reduces the mixing of the $u$ and $v$ components of the BdG spinors, implying an even lower anomalous-normal transmission.

From these results, we conclude that, apart from the quasi-stationary requirements (i) and (ii) stated in Sec. \ref{sec:themodel}, the optimal scenario for the emission of spontaneous Hawking radiation must meet the following conditions: (iii) the width of the conduction band must be lower than the threshold $\omega_{{\rm max}}$, as implied by the inequalities (\ref{eq:regimegeneral})-(\ref{eq:regimegaussian}), so the Hawking spectrum displays a highly non-thermal structure that could help to its detection; (iv) the supersonic density needs to be as high as possible in order to increase the signal of the Hawking spectrum. The latter condition is also useful for experimental purposes as it increases the signal in the measurements.

We note that there must be a trade-off between achieving a very favorable quasi-stationary regime with small fluctuations (implying a wide asymptotic conduction band with the initial chemical potential placed above its bottom) and placing the top of the conduction band below the Hawking threshold (implying that the width of the conduction band must be smaller than the threshold frequency, $\Delta_c<\omega_{\rm max}$), summarized by the conditions
\begin{equation}\label{eq:tradeoff}
\frac{E_{\rm max}}{2}\lesssim E_{\rm min}<\mu_0
\end{equation}
The above relation can be rewritten, according to Eqs. (\ref{eq:regimegeneral})-(\ref{eq:regimeideal}), as
\begin{equation}\label{eq:tradeoffideal}
\frac{2}{3}E_R\lesssim \frac{1}{2}V_{\infty}\lesssim\mu_0
\end{equation}

\subsection{Gaussian-shaped optical lattice}

\begin{figure}[!t]
\begin{tabular}{@{}cc@{}}
    \includegraphics[width=0.5\columnwidth,valign=b]{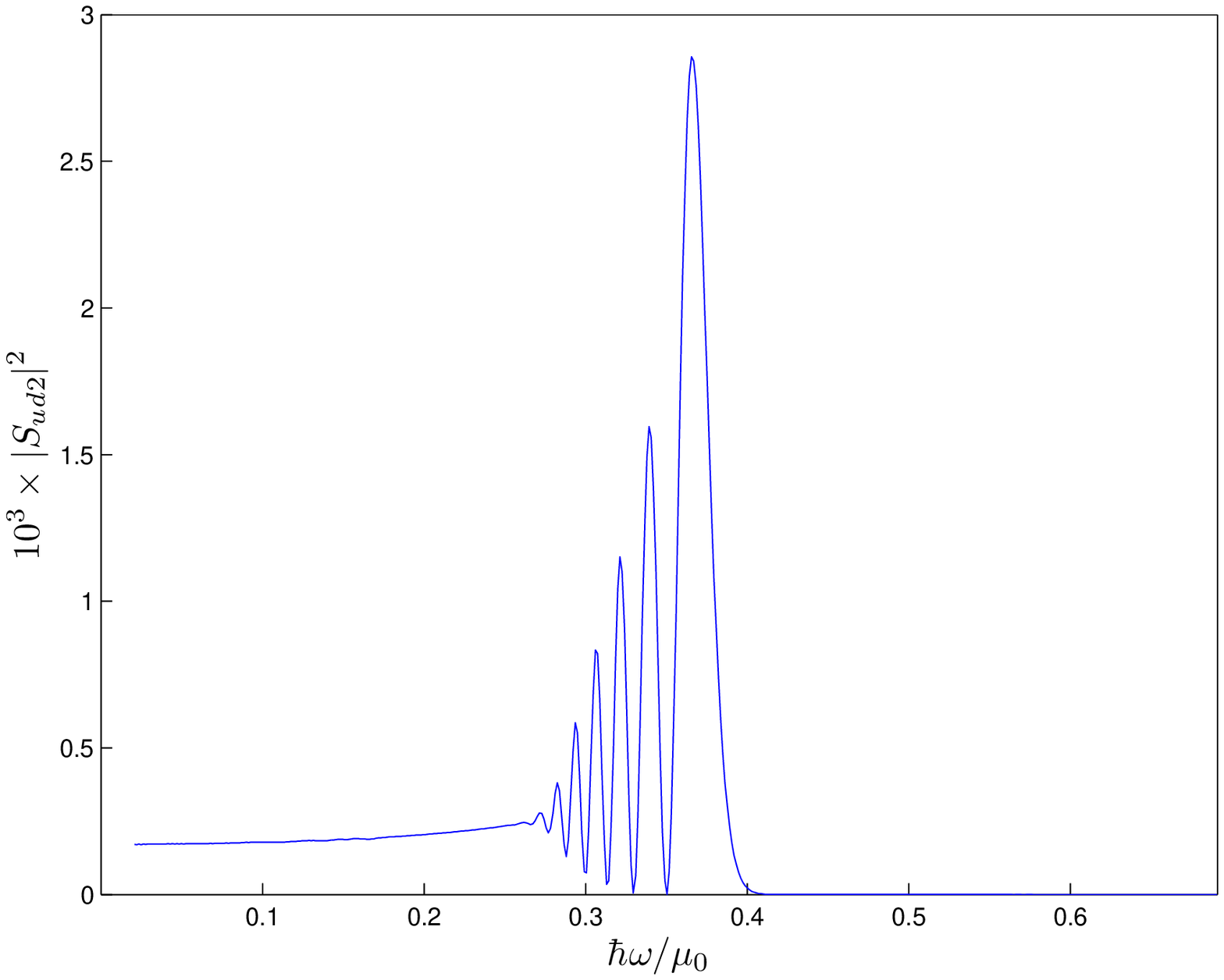} & \includegraphics[width=0.51\columnwidth,valign=b]{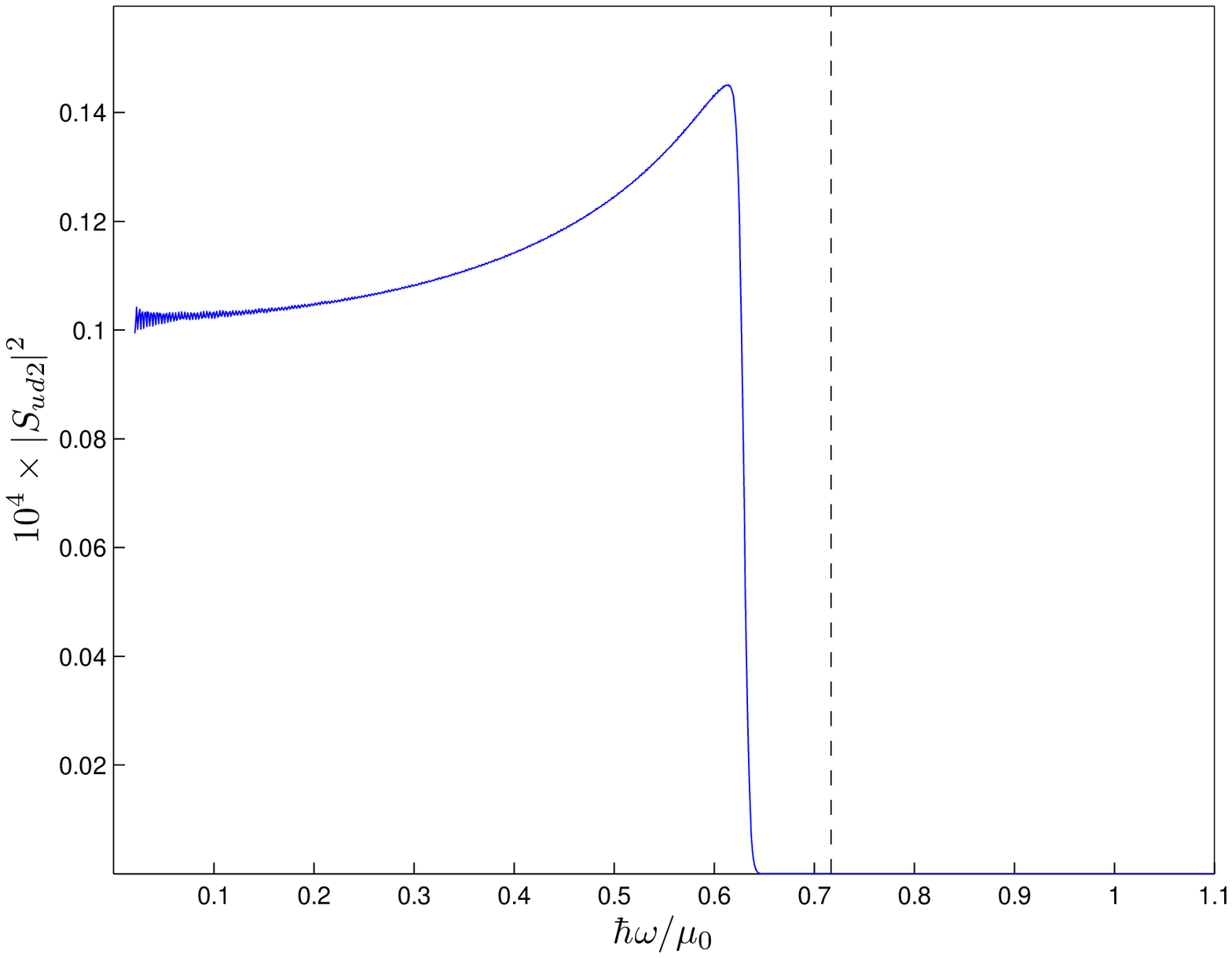}\\
    \includegraphics[width=0.5\columnwidth,valign=b]{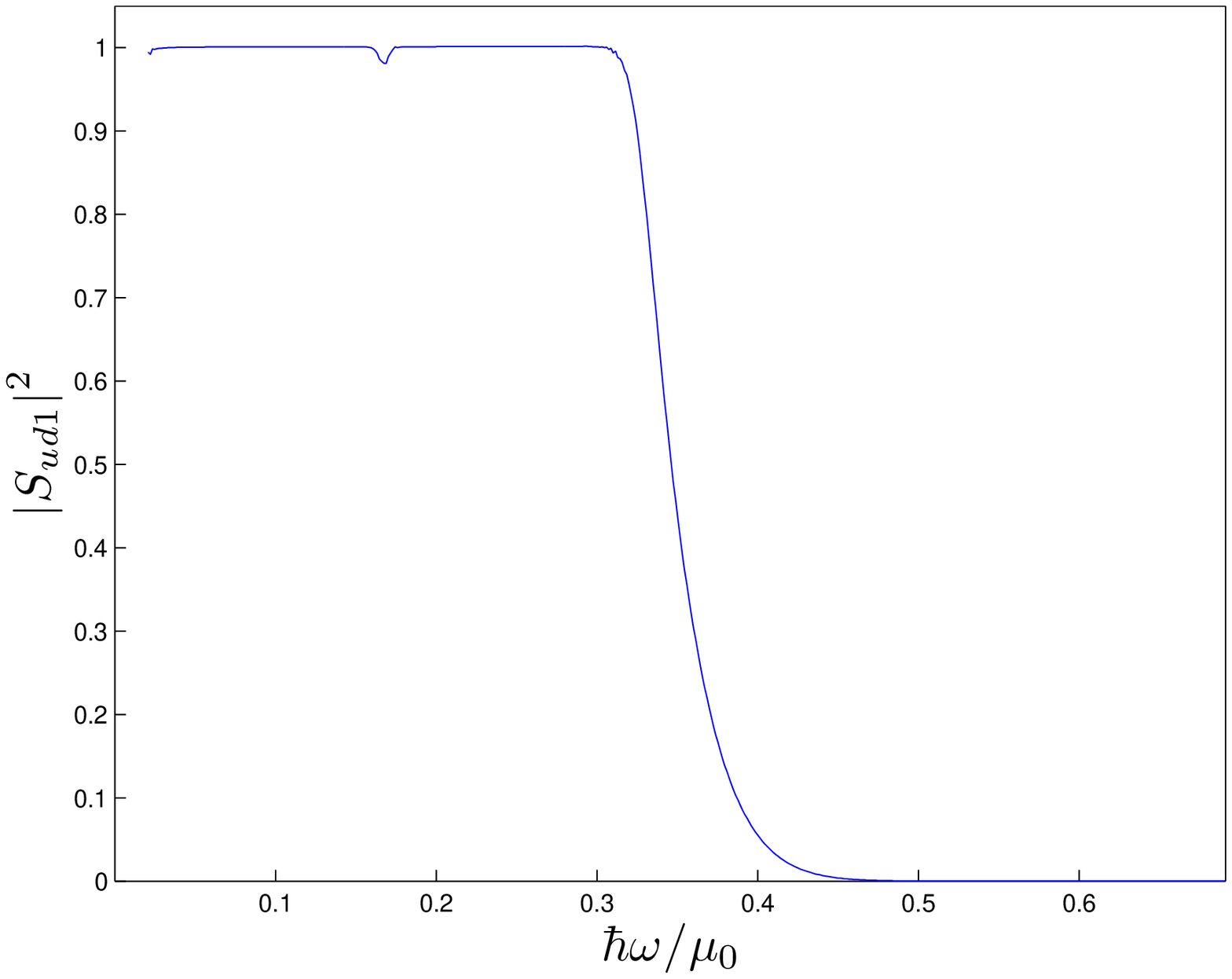} & \includegraphics[width=0.51\columnwidth,valign=b]{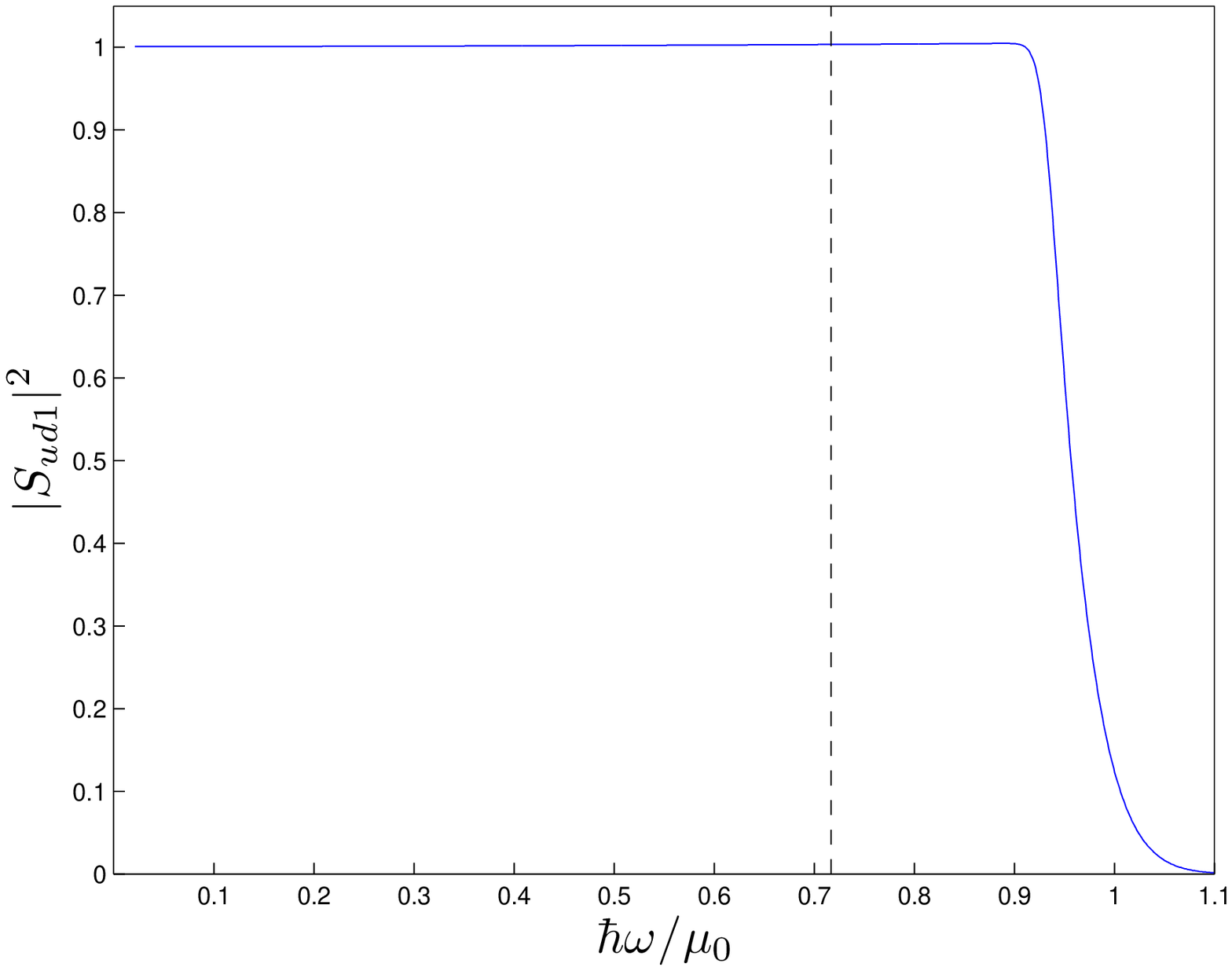}
\end{tabular}
\caption{Equivalent of Fig. \ref{fig:BdGOLIdeal} but for a realistic optical lattice with a Gaussian envelope. The snapshot of the condensate is taken at a large time $4\times 10^4\, t_{0}$. Both columns are computed for configurations $\tau=500\, t_{0}$, $L=470\,\mu\text{m}=1480~\xi_0$, $\tilde{w}=70\,\mu\text{m}=220.5~\xi_0$, $N=8.9\times 10^3$, $\omega_{\rm tr}=2\pi\times4\,\text{kHz}$ and then $n_{0}=25\,\mu\text{m}^{-1}$, $\xi_0=0.3175\,\mu\text{m}$ and $t_0=1.3792\times10^{-4}~\text{s}$. They only differ in the values of the long-time potential amplitude $V_{\infty}$ and the lattice spacing $d$. A numerical infrared cutoff $\hbar\omega_{\Lambda}=0.02\mu_0$ is introduced in both cases. Left column: $V_{\infty}=1.5~\mu_0$ and $d=2.21\xi_0$, which gives $E_{\rm min}=0.68~\mu_0$, $E_R=1.02~\mu_0$ and $\Delta_c\thickapprox E_R-E_{\rm min}=0.34~\mu_0$. The long time parameters of the quasi-stationary flow are $\bar{\mu}\simeq \omega_{\rm max}=0.69~\mu_0$, $c_u=0.83~c_0$, $v_d=1.18~c_0$ and $c_d=0.025~c_0$.
Right column: $V_{\infty}=1.5~\mu_0$ and $d=1.73\xi_0$, which gives $E_{\rm min}=0.71~\mu_0$, $E_R=1.64~\mu_0$ and $\Delta_c\thickapprox E_R-E_{\rm min}=0.93~\mu_0$. The long time parameters of the quasi-stationary flow are $\bar{\mu}\simeq \omega_{\rm max}=0.72~\mu_0$, $c_u=0.85~c_0$, $v_d=1.20~c_0$ and $c_d=0.024~c_0$. The vertical dashed line represents the threshold frequency $\omega_{\rm max}$.}
\label{fig:BdGRealistic}
\end{figure}

After using the ideal optical lattice as a test field, we now shift to the realistic optical lattice. We apply the same procedure as before and take a snapshot of the condensate at a late time $t_s=4\times 10^4\, t_{0}$, once the flow is already quasi-stationary. Most of the previous observations still hold in this realistic case. In particular, as seen in Fig. \ref{fig:BdGRealistic}, we can distinguish again between the case where the whole conduction band is placed inside the anomalous frequency range (left column) and the case where it is not (right column). In the same fashion as for the corresponding Schr\"odinger spectrum [see Fig. \ref{fig:RealisticBands}], the BdG spectrum presents a plateau in the transmission band instead of having a peaked behavior as in the ideal case. The Hawking spectrum displays strong resonant peaks near the top of the conduction band when this falls below $\omega_{\rm max}$; this regime is of special interest as it is well known that near resonant peaks one can expect a strong spontaneous Hawking signal standing above the stimulated Hawking signal \cite{Zapata2011,deNova2014,deNova2015Proc,deNova2016}. This behavior contrasts with the case where the top of the conduction band is above $\omega_{\rm max}$, in which the plateau of the transmission band only goes to zero at the end of Hawking spectrum limit without showing any remarkable structure.

For the Gaussian-enveloped potential, the trade-off condition (\ref{eq:tradeoff}) reads as
\begin{equation}\label{eq:tradeoffGaussian}
\frac{1}{2}E_R\lesssim \frac{1}{2}V_{\infty}\lesssim\mu_0\lesssim E_R
\end{equation}
where the rightmost relation arises from the quasi-stationarity requirement that $\mu_0$ must be initially placed near $E_{\rm min}$.

It is seen that, for both ideal and realistic configurations, there are strong similarities between the qualitative behavior of the Schr\"odinger and the BdG spectrum. This agreement arises due to the smallness of interactions and the low Bloch momentum of the GP wave function within the lattice, so the resulting Bloch BdG waves are close to their Schr\"odinger counterparts \cite{deNova2014a}.


\section{Conclusions and outlook} \label{sec:conclusions}

We have developed in this work an effective model to study the transport properties, including the characterization of the Hawking spectrum, of a quasi-stationary black-hole in an outcoupled Bose-Einstein through an optical lattice. For that purpose, we have numerically solved the time-independent Bogoliubov-de Gennes equations on top of the mean-field Gross-Pitaevskii wave function of the condensate, first computed in Ref. \cite{deNova2014a}. In the process, we have reviewed general concepts of Bose-Einstein condensates, optical lattices and quantum transport, and linked the numerical tools for solving the Bogoliubov-de Gennes equations with those arising in other fields of Physics such as different as propagation of ultrasonic waves or Anderson localization.

First, we have characterized the quasi-particle spectrum for a quasi-stationary black-hole in an ideal (with a flat envelope) optical lattice in order to understand the main trends. We have found
that most of the properties of the system are inherited from the linear Schr\"odinger spectrum. In particular, two qualitative regimes can be distinguished depending on whether the top of the BdG conduction band lies below or above the maximum Hawking frequency, $\omega_{\rm max}$. When the top of the conduction band is placed below $\omega_{\rm max}$, the system displays a highly peaked and structured Hawking spectrum that sharply falls to zero about the conduction band threshold, providing in this way a promising scenario for the detection of the Hawking effect due to its non-thermal character. It has been also seen that, due to the low density in the optical lattice and in the supersonic region, the normal-normal transmission greatly dominates over the anomalous-normal transmission characterizing the spontaneous Hawking emission. Consequently, it is convenient to increase the density in the supersonic region as much as possible in order to make larger the spontaneous Hawking signal as well as to increase the total experimental signal in an actual setup.

After the previous theoretical study, we have switched to the realistic case of a quasi-stationary black-hole configuration in an optical lattice with a Gaussian envelope, which can be achieved in the laboratory using standard protocols \cite{deNova2014a}. We have seen that most of the above conclusion still hold. Remarkably, as in the Schr\"odinger spectrum, both the normal-normal and anomalous-normal transmission show a plateau in the spectrum within the limits of the conduction band: for both ideal and realistic configurations, these similarities arise due to the low mean-field density and Bloch momentum inside the lattice. Regarding the spontaneous Hawking radiation, the most interesting scenario is, as in the ideal case, that in which the top of the conduction band lies below $\omega_{\rm max}$. In particular, near the top of the conduction band strong resonant peaks appear (see Fig. \ref{fig:BdGRealistic}), which are expected to provide a good signal for the characterization of the spontaneous emission of Hawking radiation \cite{Zapata2011,deNova2014,deNova2015Proc,deNova2016}. Regarding a possible experimental implementation, it is worth noting that the quasi-stationary configuration here considered is much closer to actual stationarity than the experimental setup of Ref. \cite{Steinhauer2016}, providing a cleaner background for the observation of spontaneous Hawking radiation.

We note that the scattering matrix characterizes the spontaneous Hawking emission when the state of the system is the vacuum of the incoming or outgoing states. Indeed, the temperature in experiments is sufficiently small to assume the system to be effectively at $T=0$ \cite{Steinhauer2016}, and the associated quantum state is expected to be described with some accuracy by the vacuum of the incoming states. However, a more realistic calculation should take into account that the initial quantum state of the system at $t=0$ is a thermal state of the confined quasi-particles at equilibrium at temperature $T$ and then compute its time evolution. One possible method to perform this task is the truncated Wigner method \cite{Walls2008,Sinatra2002,Carusotto2008}, in which by properly averaging over many integrations of the initial equilibrium wave function with some engineered noise on top of it, one obtains for the same price the time evolution of both the mean-field wave function and the quantum fluctuations. Nevertheless, due to the small density in the supersonic region, this method can present some problems. An alternative way could be provided by a full integration of the time-dependent BdG equations for the initial confined eigenmodes. Future works should address the computation of the time evolution of the quantum state of the system, much harder from both a theoretical and a computational point of view than that of the mean-field wave function.

Finally, we note that apart from gravitational analogue purposes, the achievement of the quasi-stationary regime here studied is of general interest in quantum transport scenarios since it provides a quasi-stationary supersonic current with very well controlled value of the velocity of the atoms. Another remarkable feature is that, as explained in the main text, the optical lattice acts as a low-pass filter of the collective modes on top of the condensate, with the band width playing the role of the cutoff frequency, which is of potential interest for the development of atomtronics. A more speculative application, but maybe feasible in some near future, is the use of the proposed black-hole configuration as a source of entangled pairs of phonons.

\acknowledgments

We thank A. Malyshev for his useful remarks about the numerical methods. We also thank D. Gu\'ery-Odelin, R. Parentani and J. Steinhauer for fruitful discussions. This work has been supported by the Israel Science Foundation, by MINECO (Spain) through grants FIS2010-21372 and FIS2013-41716-P, and by Comunidad de Madrid through grant MICROSERES-CM (S2009/TIC-1476).

\appendix

\section{General formalism} \label{app:generalformalism}

We devote this Appendix to review the general formalism of gravitational analogues in Bose-Einstein condensates. Specifically, Sec. \ref{subsec:GPBdG} discuss the mean-field approximation for condensates near zero temperature while the emergence of the gravitational analog is studied in Sec. \ref{subsec:BHBEC}.

\subsection{Gross-Pitaevskii and Bogoliubov-de Gennes equations}\label{subsec:GPBdG}

\subsubsection{Time-independent scenarios}\label{subsec:timedependent}

We begin by reviewing the mean-field approximation in Bose-Einstein condensates. For that purpose, we consider a gas of $N$ bosons near $T=0$, described by the usual second-quantization Hamiltonian \cite{Fetter2003,Dickhoff2005}:
\begin{equation}\label{eq:2ndHamiltonian}
\hat{H}=\int\mathrm{d}^3\mathbf{x}~\hat{\Psi}^{\dagger}(\mathbf{x})\left[-\frac{\hbar^2}{2m}\nabla^2+V_{\rm{ext}}(\mathbf{x})\right]\hat{\Psi}(\mathbf{x})\\
+\frac{g_{\rm{3D}}}{2}\hat{\Psi}^{\dagger}(\mathbf{x})\hat{\Psi}^{\dagger}(\mathbf{x})\hat{\Psi}(\mathbf{x})\hat{\Psi}(\mathbf{x})
\end{equation}
where $V_{\rm{ext}}$ is an external time-independent potential and the interaction between atoms is taken into account for low momentum by a contact pseudo-potential of the form $V_{\rm{int}}(\mathbf{x})=g_{\rm{3D}}\delta(\mathbf{x})$, with $g_{\rm{3D}}=4\pi\hbar^2a_s/m$ and $a_s$ the $s$-wave scattering length \cite{Pitaevskii2003,Pethick2008}. Using the canonical commutation rules
\begin{equation}\label{eq:canonical}
[\hat{\Psi}(\mathbf{x}),\hat{\Psi}^{\dagger}(\mathbf{x}')]=\delta(\mathbf{x}-\mathbf{x}')~,
\end{equation}
the equation of motion in the Heisenberg picture for the field operator $\hat{\Psi}(\mathbf{x},t)$ reads
\begin{equation}\label{eq:fieldoperatorheisenberg}
i\hbar\partial_t\hat{\Psi}(\mathbf{x},t)=[\hat{\Psi}(\mathbf{x},t),\hat{H}(t)]=\left[-\frac{\hbar^2}{2m}\nabla^2+V_{\rm{ext}}(\mathbf{x})\right]\hat{\Psi}(\mathbf{x},t)+g_{\rm{3D}}\hat{\Psi}^{\dagger}(\mathbf{x},t)\hat{\Psi}(\mathbf{x},t)\hat{\Psi}(\mathbf{x},t)
\end{equation}
Now, we make a mean-field approximation of the form
\begin{equation}\label{eq:meanfieldGP}
\hat{\Psi}(\mathbf{x},t)=[\Psi_0(\mathbf{x})+\hat{\varphi}(\mathbf{x},t)]e^{-i\frac{\mu}{\hbar}t}~,
\end{equation}
with $\Psi_0$ the expectation value of the field operator, the so-called the Gross-Pitaevskii (GP) wave function \cite{Pitaevskii2003}, and $\mu$ the chemical potential. Assuming that the population of the depletion cloud (i.e., the atoms outside the condensate) is negligible, $\Psi_0(\mathbf{x})$ is normalized to the total number of particles $N$
\begin{equation}\label{eq:normalization}
N=\int\mathrm{d}^3\mathbf{x}~|\Psi_0(\mathbf{x})|^{2}
\end{equation}

The GP wave function satisfies the {\it time-independent} GP equation
\begin{eqnarray}\label{eq:TIGP}
\nonumber H_{GP}\Psi_0(\mathbf{x})&=&\mu\Psi_0(\mathbf{x})\\
H_{GP}&=&-\frac{\hbar^2}{2m}\nabla^2+V_{\rm{ext}}(\mathbf{x})+g_{\rm{3D}}|\Psi_0(\mathbf{x})|^2
\end{eqnarray}
and describes the stationary wave function of the condensate.

On the other hand, for the quantum fluctuations of the field operator, $\hat{\varphi}(\mathbf{x},t)$, one finds to lowest order that
\begin{eqnarray}\label{eq:Fieldequation}
i\hbar\partial_t\hat{\varphi}&=&G\hat{\varphi}+L\hat{\varphi}^{\dagger},\\
\nonumber G&=&H_{GP}+g_{\rm{3D}}|\Psi_0(\mathbf{x})|^2-\mu,~L=g_{\rm{3D}}\Psi^2_0(\mathbf{x})
\end{eqnarray}
which in matrix form reads as
\begin{eqnarray}\label{eq:BdGfieldequation}
\nonumber i\hbar\partial_t\hat{\Phi}&=&M\hat{\Phi},\\
\hat{\Phi}=\left[\begin{array}{c}\hat{\varphi}\\ \hat{\varphi}^{\dagger}\end{array}\right]&,&M=\left[\begin{array}{cc}G & L\\
-L^{*}&-G\end{array}\right]
\end{eqnarray}
These equations are the well known Bogoliubov-de Gennes (BdG) equations. For illustrative purposes, we first consider their classical version, in which all the ``hats" are removed; this is also useful for practical purposes as the classical equations describe the time-evolution of the linearized collective motion of small perturbations around the stationary GP wave function, $\delta\Psi(\mathbf{x},t)$, with $\Psi(\mathbf{x},t)=[\Psi_0(\mathbf{x})+\delta\Psi(\mathbf{x},t)]e^{-i\frac{\mu}{\hbar}t}$ [see also Eqs. (\ref{eq:TDGP}), (\ref{eq:TDFieldequation})].

As usual, due to the linear character of the equations, the field fluctuations can be expanded in eigenmodes as
\begin{eqnarray}\label{eq:BdGmodesexpansion}
\nonumber\Phi(\mathbf{x},t)&=&\sum_n \gamma_nz_n(\mathbf{x})e^{-i\omega_nt}+\gamma^{*}_n\bar{z}_n(\mathbf{x})e^{i\omega_nt}\\
z_n(\mathbf{x})&=&\left[\begin{array}{c}u_n(\mathbf{x})\\ v_n(\mathbf{x})\end{array}\right],\bar{z}_n(\mathbf{x})=\left[\begin{array}{c}v^*_n(\mathbf{x})\\ u^*_n(\mathbf{x})\end{array}\right]
\end{eqnarray}
with the spinors $z_n$ satisfying the {\it time-independent} BdG equations:
\begin{equation}\label{eq:BdGequation}
Mz_n=\epsilon_nz_n, \, \epsilon_n=\hbar\omega_n
\end{equation}
This eigenvalue equation straightforwardly implies that the conjugate $\bar{z}_n$ is also a mode with eigenvalue $-\epsilon^*_n$. Interestingly, the eigenvalue problem (\ref{eq:BdGequation}) is non-Hermitian and thus it can yield {\it dynamical instabilities}, that is, exponentially growing modes (as, for instance, in a black-hole laser \cite{Finazzi2010}); we do not address this situation in the present work. Another remarkable property of Eq. (\ref{eq:BdGequation}) is that
\begin{equation}\label{eq:zeromode}
z_0=\left[\begin{array}{r}\Psi_0(\mathbf{x})\\ -\Psi^*_0(\mathbf{x})\end{array}\right]
\end{equation}
is a zero-energy mode, corresponding to the Goldstone mode arising from the gauge invariance of the Hamiltonian (\ref{eq:2ndHamiltonian}) under phase transformations $\hat{\Psi}(\mathbf{x})\rightarrow e^{i\delta}\hat{\Psi}(\mathbf{x})$.

The BdG equations present also an associated Klein-Gordon type inner product:
\begin{equation}\label{eq:KGProduct}
(z_m|z_n)\equiv\braket{z_m|\sigma_z|z_n}=\int\mathrm{d}^3\mathbf{x}~z_m^{\dagger}(\mathbf{x})\sigma_zz_n(\mathbf{x})=\int\mathrm{d}^3\mathbf{x}~u^*_m(\mathbf{x})u_n(\mathbf{x})-v^*_m(\mathbf{x})v_n(\mathbf{x}) \, ,
\end{equation}
with $\sigma_z=\rm{diag}(1,-1)$ the corresponding Pauli matrix, which presents the following property for eigenmodes
\begin{equation}\label{eq:orthogonal}
(\epsilon_n-\epsilon^*_m)(z_m|z_n)=0
\end{equation}
Then, as usual, two modes with different eigenvalues are orthogonal according to this scalar product. However, it is worth noting that it is not positive definite so the norm of a given solution $z$, defined as $(z|z)$, can be positive, negative or zero. Indeed, the norm of $\bar{z}$ has the opposite sign to that of $z$, $(z|z)=-(\bar{z}|\bar{z})$; in particular, the zero-energy mode of Eq. (\ref{eq:zeromode}) has zero norm and $\bar{z_0}=-z_0$. The orthogonality relation (\ref{eq:orthogonal}) also implies that complex modes have zero norm. In the following, we only consider normalized modes with norm $\pm 1$.

Turning back to the quantum case, the amplitude of each mode is promoted to a quantum operator $\gamma_n\rightarrow\hat{\gamma}_n$ as
\begin{equation}\label{eq:destructoroperators}
\hat{\gamma}_n=(z_n|\hat{\Phi}) \, ,
\end{equation}
which implies, via canonical commutation rules (\ref{eq:canonical}), that $[\hat{\gamma}_n,\hat{\gamma}_{n'}]=-(z_n|\bar{z}_{n'})=0$ and $[\hat{\gamma}_n,\hat{\gamma}^{\dagger}_{n'}]=(z_n,z_{n'})=\pm\delta_{nn'}$ where the $\pm$ sign depends on whether the mode has positive (negative) norm; we do not discuss here the quantization of modes with zero norm (the interested reader can check Ref. \cite{Castin1998} for the quantization of the zero-energy mode or Ref. \cite{Finazzi2010} for the case of complex modes in the context of the black-hole laser). Hence, the amplitude of modes with positive (negative) norm corresponds to an annihilation (creation) operator. In particular, as the norm of the conjugate $\bar{z}_n$ has opposite sign to that of $z_n$, its associated amplitude has the opposite character to that of $z_n$. From now on, $\hat{\gamma}_n$ only denotes annihilation operators.

A different approach to the BdG equations arises from considering the grand-canonical Hamiltonian $\hat{K}=\hat{H}-\mu\hat{N}$, with $\hat{H}$ given by Eq. (\ref{eq:2ndHamiltonian}) and $\hat{N}$ the particle number operator:
\begin{equation}\label{eq:numberparticles}
\hat{N}=\int\mathrm{d}^3\mathbf{x}~\hat{\Psi}^{\dagger}(\mathbf{x})\hat{\Psi}(\mathbf{x})
\end{equation}
Using the mean-field approximation (\ref{eq:meanfieldGP}) and keeping consistently only up to quadratic terms in $\hat{\varphi}$, we find that the grand-canonical Hamiltonian can be decomposed as
\begin{equation}\label{eq:BogoliubovDecompositionK}
\hat{K}=K[\Psi_0]+K_V+\hat{K}_B
\end{equation}
The term $K[\Psi_0]$ is the mean-field contribution, resulting from the substitution of $\hat{\Psi}$ by $\Psi_0$ in the expression for $\hat{K}$, with
\begin{equation}\label{eq:Kfunctional}
K[\Psi]=\int\mathrm{d}^3\mathbf{x}~\Psi^{*}(\mathbf{x})\left[-\frac{\hbar^2}{2m}\nabla^2+V_{\rm{ext}}(\mathbf{x})\right]\Psi(\mathbf{x})+\frac{g_{\rm{3D}}}{2}|\Psi(\mathbf{x})|^4-\mu|\Psi(\mathbf{x})|^2
\end{equation}
As $\Psi_0(\mathbf{x})$ is a solution of Eq. (\ref{eq:TIGP}), $K[\Psi_0]$ can be evaluated as
\begin{equation}\label{eq:meanfieldhamiltonian}
K[\Psi_0]=-\int\mathrm{d}^3\mathbf{x}~\frac{g_{\rm{3D}}}{2}|\Psi_0(\mathbf{x})|^4
\end{equation}
Secondly, the contribution from the depletion cloud, $K_V$, is:
\begin{equation}\label{eq:depletionhamiltonian}
K_V=-\sum_n\int\mathrm{d}^3\mathbf{x}~\epsilon_n|v_n(\mathbf{x})|^2
\end{equation}
and arises from the non-commutative character of the annihilation and creation operators of the quasi-particles (the eigenmodes of the BdG equations). Finally, the proper quantum contribution is contained in the Bogoliubov Hamiltonian
\begin{equation}\label{eq:Bogoliubovhamiltonian}
\hat{K}_B=\sum_n\epsilon_n\hat{\gamma}^{\dagger}_{n}\hat{\gamma}_n
\end{equation}
Hence, in this approximation, called the Bogoliubov approximation, the grand-canonical Hamiltonian is diagonalized by the solutions of the BdG equations. We note that there is no linear term in the $\hat{\gamma}_n$ operators as the time-independent GP equation (\ref{eq:TIGP}) is precisely the condition for $\Psi_0$ to be an extreme of the mean-field grand canonical functional $K[\Psi]$. If $\Psi_0$ is a minimum of this functional, it is said that the solution is energetically, statically or Landau stable. On the other hand, if it is an extreme but not a local minimum, $\Psi_0$ is energetically unstable since any perturbation would induce the system to decay to a lower energy state. In order to further check the energetic stability of the solution, one can consider linear perturbations around the stationary solution, $\Psi=\Psi_0+\delta \Psi$, finding:
\begin{eqnarray}\label{eq:Landauperturbation}
\nonumber \delta K&=&K[\Psi]-K[\Psi_0]=\int\mathrm{d}^3\mathbf{x}~\frac{1}{2}\delta\Phi^{\dagger}\Lambda\delta\Phi\\
\delta\Phi&=&\left[\begin{array}{c}\delta \Psi\\ \delta \Psi^{*}\end{array}\right],~\Lambda=\sigma_z M=\left[\begin{array}{cc}G & L\\
L^{*}&G\end{array}\right]
\end{eqnarray}
Note the strong similarity with the expansion of Eq. (\ref{eq:BogoliubovDecompositionK}); in fact, the expression for $\delta K$ is the classical version of the Bogoliubov Hamiltonian (\ref{eq:Bogoliubovhamiltonian}) provided that $\Psi_0$ is a local extreme as $K_V$ only appears in the quantum version due to the non-commutativity of the quantum operators. The eigenvalues of the operator $\Lambda$, given by a similar equation to the eigenvalue BdG equation (\ref{eq:BdGequation}),
\begin{equation}
\Lambda\left[\begin{array}{c} u \\
v\end{array}\right]=\lambda\left[\begin{array}{c} u \\
v\end{array}\right],~
\end{equation}
characterize the stability of the solution. If all of them are positive (except the zero-energy Goldstone mode, still present), the state $\Psi_0$ is energetically stable and in the opposite case the state is energetically unstable. In particular, as argued by Landau and discussed later, supersonic flows are energetically unstable. Note that $\Lambda$ is an Hermitian matrix operator and it only presents real eigenvalues, in contrast to the operator $M$ characterizing the BdG equations. The close link between Landau stability and BdG equations is revealed by the relation
\begin{equation}
\braket{z_n|\Lambda|z_n}=\epsilon_n (z_n|z_n)
\end{equation}
which shows that, if the state is Landau stable, there are no complex modes since then $(z_n|z_n)=0$. Also, for a Landau stable state, a mode with positive (negative) energy has positive (negative) normalization. Thus, the presence of an {\it anomalous} mode, i.e., a mode with positive (negative) normalization and negative (positive) energy, implies that the system is energetically unstable. 

In a more physical picture, the minimization of $K$ can be understood as the minimization of the Hamiltonian $H$ for fixed number of particles $N$, with the chemical potential $\mu$ playing the role of the Lagrange multiplier.

\subsubsection{Time-dependent scenarios}\label{subsec:timedependent}

The previous considerations can be extended to time-dependent scenarios by writing the field operator in Eq. (\ref{eq:Fieldequation}) in a more general mean-field approximation, $\hat{\Psi}(\mathbf{x},t)=\Psi(\mathbf{x},t)+\hat{\varphi}(\mathbf{x},t)$. As a result, Eq. (\ref{eq:TIGP}) is extended to the {\it time-dependent} GP equation:
\begin{eqnarray}\label{eq:TDGP}
H_{GP}(t)\Psi(\mathbf{x},t)&=&i\hbar\partial_t\Psi(\mathbf{x},t)\\
\nonumber  H_{GP}(t)&=&-\frac{\hbar^2}{2m}\nabla^2+V_{\rm{ext}}(\mathbf{x},t)+g_{\rm{3D}}|\Psi(\mathbf{x},t)|^2
\end{eqnarray}
where we have allowed for a possible time-dependence of the external potential. The normalization of the GP wave function is kept constant as guaranteed by the continuity equation
\begin{equation}\label{eq:GPnormconservation}
\partial_t|\Psi(\mathbf{x},t)|^2+\nabla\mathbf{J}(\mathbf{x},t)=0,~\mathbf{J}(\mathbf{x},t)=-\frac{i\hbar}{2m}\left[\Psi^*(\mathbf{x},t)\nabla \Psi(\mathbf{x},t)-\Psi(\mathbf{x},t)\nabla \Psi^*(\mathbf{x},t)\right]
\end{equation}
with $\mathbf{J}(\mathbf{x},t)$ the current. It is instructive to rewrite these results in terms of the amplitude and phase of the wave function, $\Psi(\mathbf{x},t)=A(\mathbf{x},t)e^{i\theta(\mathbf{x},t)}$,
\begin{eqnarray}\label{eq:PhaseAmplitude}
\partial_tn+\nabla(n\mathbf{v})&=&0\\
\nonumber -\hbar \partial_t \theta&=&-\frac{\hbar^2}{2mA}\nabla^2A+\frac{1}{2}m\mathbf{v}^2+V_{\rm{ext}}(\mathbf{x},t)+g_{\rm{3D}}A^2 \\
\nonumber n(\mathbf{x},t)&\equiv&A^2(\mathbf{x},t),~\mathbf{v}(\mathbf{x},t)\equiv\frac{\hbar\nabla\theta(\mathbf{x},t)}{m}~;
\end{eqnarray}
$n(\mathbf{x},t),\mathbf{v}(\mathbf{x},t)$ being the local mean-field density and flow velocity, respectively. Interestingly, the first line of Eq. (\ref{eq:PhaseAmplitude}) is the same conservation law of Eq. (\ref{eq:GPnormconservation}) as $\mathbf{J}=n\mathbf{v}$ and it is the equivalent of the continuity equation for a hydrodynamical fluid, motivating the chosen name for Eq. (\ref{eq:GPnormconservation}). Moreover, the second line can be also rewritten as the analog of the Euler equation for the velocity of a potential flow by taking the gradient on both sides of the equation:
\begin{equation}\label{eq:Hydrodynamics}
m\partial_t \mathbf{v}=-\nabla\left[\frac{1}{2}m\mathbf{v}^2+V_{\rm{ext}}(\mathbf{x},t)\right]-\frac{1}{n}\nabla P(\mathbf{x},t)+\nabla\left[\frac{\hbar^2}{2m\sqrt{n}}\nabla^2\sqrt{n}\right],~P(\mathbf{x},t)=\frac{gn^2(\mathbf{x},t)}{2}
\end{equation}
where $P$ is just the local pressure of the condensate, as given by the equation of state of an uniform condensate at equilibrium, $P=\frac{gn^2}{2}$. The only difference with the usual hydrodynamic equation is the rightmost term, often called the {\em quantum pressure} term, representing a genuine quantum feature as it contains $\hbar$.

It is very useful to rederive the above equations within a Lagrangian frame; it is straightforward to prove that the time-dependent GP equation (\ref{eq:TDGP}) results from the equations of motion of the following Lagragian:
\begin{equation}\label{eq:Lagrangian}
\mathfrak{L}[\Psi]=\int\mathrm{d}^3\mathbf{x}\mathrm{d}t~i\hbar\Psi^{*}(\mathbf{x},t)\partial_t \Psi(\mathbf{x},t)
-\frac{\hbar^2}{2m}|\nabla\Psi(\mathbf{x},t)|^2-V_{\rm{ext}}(\mathbf{x},t)|\Psi(\mathbf{x},t)|^2-\frac{g_{\rm{3D}}}{2}|\Psi(\mathbf{x},t)|^4
\end{equation}

With respect to the quantum fluctuations, their time evolution is given by:
\begin{eqnarray}\label{eq:TDFieldequation}
\nonumber i\hbar\partial_t\hat{\Phi}&=&M(t)\hat{\Phi},\\
M(t)&=&\left[\begin{array}{cc}G(t) & L(t)\\
-L^{*}(t)&-G(t)\end{array}\right]\\
\nonumber G(t)&=&H_{GP}(t)+g_{\rm{3D}}|\Psi(\mathbf{x},t)|^2\\
\nonumber L(t)&=&g_{\rm{3D}}\Psi^2(\mathbf{x},t)
\end{eqnarray}
A possible way to integrate the previous equation for the quantum fluctuations of the field operator is to decompose it in terms of a complete set of solutions at the initial moment and solve separately the resulting BdG equations for each mode $z_n(t)$:
\begin{equation}\label{eq:TDBdG}
M(t)z_n(t)=i\hbar\partial_t z_n(t),
\end{equation}
The scalar product (\ref{eq:KGProduct}) of two different modes $z_1(t),z_2(t)$ is preserved during the time evolution by the previous equation; consequently, the norm is also a conserved quantity, with an associated quasiparticle current given by:
\begin{equation}\label{eq:quasiparticlecurrent}
\mathbf{J}_z=-\frac{i\hbar}{2m}\left(u^*\nabla u-u\nabla u^*+v^*\nabla v-v\nabla v^*\right)
\end{equation}
where $u(\mathbf{x},t),v(\mathbf{x},t)$ are the components of a given spinor solution $z(\mathbf{x},t)$. The corresponding continuity equation reads:
\begin{equation}\label{eq:BdGnormconservation}
\partial_t(z^{\dagger}\sigma_zz)+\nabla\mathbf{J}_z=0
\end{equation}
As natural, for stationary GP solutions $\Psi(\mathbf{x},t)=\Psi_0(\mathbf{x})e^{-i\frac{\mu}{\hbar}t}$, the {\it time-dependent} GP equation (\ref{eq:TDGP}) is reduced to the {\it time-independent} GP equation (\ref{eq:TIGP}). In addition, by removing the phase $e^{-i\frac{\mu}{\hbar}t}$ from the operator $\hat{\varphi}$, the {\it time-dependent} BdG equations (\ref{eq:TDFieldequation}) are also transformed into the stationary BdG equations (\ref{eq:BdGfieldequation}). In this stationary regime, the conservation laws (\ref{eq:GPnormconservation}), (\ref{eq:BdGnormconservation}) simply read as:

\begin{equation}\label{eq:stationarycurrentssolutions}
\nabla\mathbf{J}=0,~\nabla\mathbf{J}_z=0
\end{equation}

\subsubsection{Effective 1D regime}\label{subsec:1Dmeanfield}

In order to describe an effective one-dimensional configuration along the $x$-axis, we consider an external potential of the form $V_{\rm{ext}}(\mathbf{x},t)=V(x,t)+\frac{1}{2}m\omega_{\rm{tr}}^2\rho^2$, where the first term $V(x,t)$ represents an external potential that only depends on the $x$ coordinate while the second term corresponds to a transverse harmonic confinement, very typical in experimental setups, $\rho=\sqrt{y^2+z^2}$ being the radial distance to the $x$-axis. After inserting the following ansatz for the wave function
\begin{equation}
\Psi(\mathbf{x},t)=\psi(x,t)\frac{e^{-\frac{\rho^2}{2\sigma^2(x,t)}}}{\sqrt{\pi}\sigma(x,t)}
\end{equation}
in the Lagrangian (\ref{eq:Lagrangian}) and integrating along the transverse degrees of freedom, it is found that \cite{Salasnich2002}
\begin{eqnarray}\label{eq:LagrangianEffective}
\nonumber \mathfrak{L}[\psi,\sigma]&=&\int\mathrm{d}x\mathrm{d}t~i\hbar\psi^{*}(x,t)\partial_t \psi(x,t)
-\frac{\hbar^2}{2m}|\partial_x\psi(x,t)|^2
-\left[V(x,t)+\frac{\hbar^2}{2m\sigma^2(x,t)}+\frac{1}{2}m\omega_{\rm{tr}}^2\sigma^2(x,t)\right]|\psi(x,t)|^2\\
&-&\frac{g_{\rm{3D}}}{4\pi\sigma^2(x,t)}|\psi(x,t)|^4
\end{eqnarray}
where we have assumed that the length scale of variation of $\sigma(x,t)$ is sufficiently large to neglect the corresponding derivative. The corresponding equations of motion for $\psi(x,t)$ and $\sigma(x,t)$ read:
\begin{eqnarray}\label{eq:Lagrangianequationsofmotion}
\nonumber i\hbar\partial_t\psi(x,t)&=&-\frac{\hbar^2}{2m}\partial^2_x\psi(x,t)
+\left[V(x,t)+\frac{\hbar^2}{2m\sigma^2(x,t)}+\frac{1}{2}m\omega_{\rm{tr}}^2\sigma^2(x,t)
+\frac{g_{\rm{3D}}}{2\pi\sigma^2(x,t)}|\psi(x,t)|^2\right]\psi(x,t)\\
0&=&\left[1+2|\psi(x,t)|^2a_s\right]\frac{\hbar^2}{m\sigma^3(x,t)}-m\omega_{\rm{tr}}^2\sigma(x,t)
\end{eqnarray}
from which we obtain:
\begin{equation}
\sigma(x,t)=a_{\rm{tr}}\left[1+2|\psi(x,t)|^2a_s\right]^{\frac{1}{4}},~a_{\rm{tr}}=\sqrt{\frac{\hbar}{m\omega_{\rm{tr}}}}~,
\end{equation}
$a_{\rm{tr}}$ being the transverse harmonic oscillator length. By plugging the expression for $\sigma(x,t)$ into the equation for $\psi(x,t)$ we finally arrive at the effective non-polynomial Gross-Pitaevskii (NPGP) equation for the 1D wave function \cite{Salasnich2002}
\begin{eqnarray}\label{eq:NPGPE}
\nonumber i\hbar\partial_t\psi(x,t)&=&-\frac{\hbar^2}{2m}\partial^2_x\psi(x,t)
+V(x,t)\psi(x,t)+\frac{g_{\rm{1D}}}{\sqrt{1+2|\psi(x,t)|^2a_s}}|\psi(x,t)|^2\psi(x,t)\\
\nonumber &+&\frac{\hbar\omega_{\rm{tr}}}{2}\left[\sqrt{1+2|\psi(x,t)|^2a_s}+\frac{1}{\sqrt{1+2|\psi(x,t)|^2a_s}}\right]\psi(x,t)\\
g_{\rm{1D}}&\equiv&\frac{g_{\rm{3D}}}{2\pi a^2_{\rm{tr}}}=2\hbar\omega_{\rm{tr}}a_s~.
\end{eqnarray}
The previous equation has been shown to describe quite well the effective 1D dynamics of a trapped condensate in a wide range of situations \cite{Salasnich2002,Tettamanti2016}. Due to the chosen normalization, $n(x,t)=|\psi(x,t)|^2$ is the 1D density
\begin{equation}
n(x,t)=\int\mathrm{d}y\mathrm{d}z~n(\mathbf{x},t)=\int\mathrm{d}y\mathrm{d}z~|\Psi(\mathbf{x},t)|^2
\end{equation}
and then $\psi(x,t)$ is normalized to the total number of particles $N$, satisfying also the corresponding 1D continuity equation, $\partial_t n(x,t)+\partial_x [n(x,t)v(x,t)]=0$.

A particular case of special interest is the low-density limit
\begin{equation}\label{eq:1Dmeanfieldregime}
n(x,t)a_s\ll1
\end{equation}
in which the previous NPGP equation is reduced (after absorbing the zero-point energy of the harmonic oscillator $\hbar\omega_{\rm{tr}}$ in the Hamiltonian) to the usual 1D GP equation
\begin{eqnarray}\label{eq:TIGP1D}
\nonumber H_{GP\rm{1D}}(t)\psi(x,t)&=&i\hbar\partial_t\psi(x,t)\\
H_{GP\rm{1D}}(t)&=&-\frac{\hbar^2}{2m}\partial_x^2+V(x,t)+g_{\rm{1D}}|\psi(x,t)|^2
\end{eqnarray}
in a regime known as the 1D mean-field regime \cite{Leboeuf2001,Menotti2002}. In this regime, one can also derive along the same lines an effective 1D BdG equation for the quantum fluctuations:
\begin{eqnarray}\label{eq:BdGfieldequation1D}
\nonumber i\hbar\partial_t\hat{\Phi}_{1D}&=&M_{\rm{1D}}(t)\hat{\Phi}_{1D},\\
M_{\rm{1D}}(t)&=&\left[\begin{array}{cc}G_{\rm{1D}}(t) & L_{\rm{1D}}(t)\\
-L_{\rm{1D}}^{*}(t)&-G_{\rm{1D}}(t)\end{array}\right]\\
\nonumber G_{\rm{1D}}(t)&=&H_{GP\rm{1D}}(t)+g_{\rm{1D}}|\psi_0(x,t)|^2,~L_{\rm{1D}}(t)=g_{\rm{1D}}\psi^2_0(x,t)
\end{eqnarray}
and perform a mode expansion for $\hat{\Phi}_{1D}$ similar to that of Eq. (\ref{eq:BdGmodesexpansion}), obtaining the 1D version of the BdG equations (\ref{eq:TDBdG}). This 1D Bogoliubov approximation is expected to be valid in the regime
\begin{equation}\label{eq:Tonkscondition}
na^2_{\rm{tr}}/a_s\gg1
\end{equation}
since otherwise the system behaves as a Tonks-Girardeau gas \cite{Menotti2002,Dunjko2001}. In the rest of the work, we drop the index $\rm{1D}$ as we will restrict ourselves to 1D configurations in the 1D mean-field regime (\ref{eq:TIGP1D}). We also identify $\Psi(x,t)$ with $\psi(x,t)$, since the motion of the transverse degrees of freedom is frozen.

\subsection{Black holes in Bose-Einstein condensates}\label{subsec:BHBEC}

\subsubsection{Scattering through a subsonic-supersonic interface}

After showing the procedure to reach an effective 1D configuration, we study in this section the stationary flow of a 1D condensate within the 1D mean-field regime, in which the gravitational analogy emerges more naturally. For that purpose, we begin by considering homogeneous flows, characterized by plane wave solutions for the GP wave function, $\Psi_0(x)=\sqrt{n}e^{iqx+\theta_0}$. The corresponding BdG modes also read in terms of plane waves with wave vector $k$ and energy $\epsilon=\hbar\omega$, connected through the dispersion relation:
\begin{equation}\label{eq:dispersionrelation}
\left[\omega-vk\right]^{2}=\Omega^2=c^{2}k^{2}+\frac{\hbar^2k^{4}}{4m^2}
\end{equation}
with $c=\sqrt{gn/m}$ the sound velocity, $v=\hbar q/m$ the constant flow velocity and $\Omega$ the comoving frequency. The healing length is defined here as $\xi\equiv\hbar/mc$. The above dispersion relation is nothing else than the usual Bogoliubov dispersion relation for phonons in a condensate at rest, $\Omega(k)$, combined with the Doppler shift due to the fluid velocity $v$. For convention, we always consider the flow velocity and comoving frequency as positive $v,\Omega>0$. According to the defined magnitudes, the system is denoted as supersonic when $v>c$ and subsonic when $v<c$.

The dispersion relation (\ref{eq:dispersionrelation}) presents four solutions for $k$ for a given laboratory frequency $\omega$; the sum and the product of these four wave vectors, labeled with the index $a$, satisfy
\begin{equation}\label{eq:wavevectorsum}
\sum_{a}k_a(\omega)=0,~\Pi_{a} k_a(\omega)=-\frac{4m^2\omega^2}{\hbar^2}
\end{equation}
The expression for the spinor solution of the BdG equations for each wave vector $k_a$ is
\begin{eqnarray} \label{eq:PlaneWaveSpinors}
s_{a,\omega}\left(x\right) & \equiv & \frac{e^{ik_{a}\left(\omega\right)x}}{\sqrt{2\pi|w_{a}\left(\omega\right)|}}\left[\begin{array}{c}
e^{i(qx+\theta_0)}u_{a}(\omega)\\
e^{-i(qx+\theta_0)}v_{a}(\omega)
\end{array}\right] \nonumber\\
\left[\begin{array}{c}
\nonumber u_{a}(\omega)\\
v_{a}(\omega)
\end{array}\right]&=&N_a\left[\begin{array}{c}
\frac{\hbar k_{a}^{2}\left(\omega\right)}{2m}+[\omega-vk_{a}\left(\omega\right)]\\
\frac{\hbar k_{a}^{2}\left(\omega\right)}{2m}-[\omega-vk_{a}\left(\omega\right)]
\end{array}\right]\\
N_a&=&\left(\frac{m}
{2\hbar k_{a}^{2}\left(\omega\right)\left|\omega-vk_{a}\left(\omega\right)\right|}\right)^{\frac{1}{2}}.
\end{eqnarray}
with $w_{a}\left(\omega\right)\equiv\left[dk_{a}\left(\omega\right)/d\omega\right]^{-1}$ the group velocity of each mode; it is included in the definition of the solutions in order to normalize the propagating modes (those with purely real wave vector) in frequency domain, $(s_{a,\omega}|s_{a,\omega'})=\pm\delta\left(\omega-\omega'\right)$. We note that all normalization factors can be removed for modes with complex wave vectors as in that case they do not play any special role.

The dispersion relation for a subsonic (supersonic) flow is represented in left (right) Fig. \ref{fig:DispRelation} of the main text. In particular, after rewriting Eq. (\ref{eq:dispersionrelation}) as $\omega=vk\pm \Omega$, the blue (red) curves represent the sign $+(-)$ branches; they also correspond to positive (negative) norm of the associated BdG modes. For subsonic flows, there are only two propagating modes for a given value of the frequency $\omega>0$, one with positive group velocity and the other with negative one, both modes presenting positive normalization. The other two solutions have complex wave vector, being one the complex conjugate of the other so one is exponentially increasing and the other exponentially decreasing. For supersonic flows, in the window $0<\omega<\omega_{\rm max}$, all the four modes are propagating, structured into two pairs of modes with positive (negative) normalization, labeled as $d1(d2)$; within each pair, as for subsonic flows, each mode has positive/negative group velocity. The frequency $\omega_{\rm max}$ is the threshold frequency as it represents the upper limit of the spectrum of Hawking radiation, as explained later. Above $\omega_{\rm max}$, the anomalous $d2$ modes are no longer propagating and thus, there is no Hawking effect. The presence of such anomalous modes arises from the Landau instability of supersonic flows.

The previous magnitudes can be extended to non-homogeneous configurations by taking $c(x)\equiv\sqrt{gn(x)/m}$ and $v(x)$ as defined in Eq. (\ref{eq:PhaseAmplitude}). In an analog way, the condensate is subsonic where $v(x)<c(x)$ and supersonic where $v(x)>c(x)$. A black-hole configuration is defined as that with two asymptotic homogeneous regions, one subsonic and one supersonic, with the flow going from subsonic to supersonic. On the other hand, if the flow travels from supersonic to subsonic, we have a white-hole configuration, the time reversal of a black-hole one (simply obtained after conjugating the wave function). By continuity, a black-hole configuration implies that, at least, one sonic horizon is formed, that is, a point where $v(x)=c(x)$. We denote the region between the two asymptotic regions, where the sonic horizon is placed, as the scattering region.


Within the convention chosen in this work, where the flow velocity is taken as positive, the {\it upstream} subsonic region (labeled as ``u") is placed at $x\rightarrow-\infty$ while the {\it downstream} supersonic region (labeled as ``d") is located at $x\rightarrow\infty$, matching in this way the notation ``u" and ``d" of Fig. \ref{fig:DispRelation}. The different asymptotic modes can be further classified according to their group velocity. Specifically, in the subsonic region, traveling waves with positive or negative group velocity correspond to incoming (traveling towards the horizon) or outgoing (traveling outward the horizon) waves, respectively. In the supersonic region, the situation is the opposite and modes with positive (negative) group velocity are outgoing (incoming) modes. For simplicity, we label the incoming modes as ``in" and the outgoing modes as ``out".

The analogy with astrophysical black holes arises when considering acoustic phonons, corresponding to modes with vanishing wave vector, since they are trapped in the supersonic side of a sonic horizon, dragged away by the flow, as light is trapped in a black hole inside the event horizon. However, due to the superluminal comoving dispersion relation $\Omega(k)$, modes with sufficiently large wave vector in the supersonic region (corresponding to incoming modes) can still travel upstream and escape from the acoustic black hole unlike in gravitational black holes where nothing escapes.

For given frequency $\omega$, the global stationary solutions to the BdG equations in a black-hole configuration can be written, in the asymptotic homogeneous regions, as combinations of the different plane wave solutions (scattering channels). Specifically, the retarded (``in") scattering states $z_{a,\omega}(x)\equiv\left[u_{a,\omega}(x),v_{a,\omega}(x)\right]^{T}$ are those modes with unit amplitude in the incoming channel $a=d1,d2,u-{\rm in}$ and zero amplitude in the remaining ones. The amplitude of the ``out" scattering channels for these scattering solutions are given by the $S$-matrix, as usual in scattering theory. For instance, the asymptotic expression for the $d2$-in mode reads:
\begin{eqnarray}\label{eq:scatteringchannelstate}
z_{d2-\rm{in},\omega}\left(x\rightarrow-\infty\right)&=&S_{ud2}\left(\omega\right)s_{u-\rm{out},\omega}\left(x\right)\\
\nonumber z_{d2-\rm{in},\omega}\left(x\rightarrow\infty\right)&=&s_{d2-\rm{in},\omega}\left(x\right)+S_{d1d2}\left(\omega\right)s_{d1-\rm{out},\omega}\left(x\right)
+S_{d2d2}\left(\omega\right)s_{d2-\rm{out},\omega}\left(x\right)
\end{eqnarray}
The remaining ``in" scattering states can be written in a similar fashion. On the other hand, the advanced (``out") scattering states are the analog of the ``in" states but changing the ``in" by ``out", i.e., they have unit amplitude in one outgoing channel and zero in the other outgoing channels.

Globally, the ``out" and ``in" states are related through the scattering matrix as:
\begin{equation}\label{eq:inoutmodesrelationspinors}
z_{i-\rm{in},\omega}=\sum_{j}S_{ji}(\omega)z_{j-\rm{out},\omega}
\end{equation}
which implies the following relation for the corresponding quantum operators
\begin{equation}\label{eq:inoutmodesrelation}
\left[\begin{array}{c}
\hat{\gamma}_{u-\rm{out}}\\
\hat{\gamma}_{d1-\rm{out}}\\
\hat{\gamma}_{d2-\rm{out}}^{\dagger}
\end{array}\right] = S(\omega)\left[\begin{array}{c}
\hat{\gamma}_{u-\rm{in}}\\
\hat{\gamma}_{d1-\rm{in}}\\
\hat{\gamma}_{d2-\rm{in}}^{\dagger}
\end{array}\right],~S(\omega)\equiv\left[\begin{array}{ccc}S_{uu}(\omega)&S_{ud1}(\omega)&S_{ud2}(\omega)\\
S_{d1u}(\omega)&S_{d1d1}(\omega)&S_{d1d2}(\omega)\\
S_{d2u}(\omega)&S_{d2d1}(\omega)&S_{d2d2}(\omega)\end{array}\right],
\end{equation}
$\hat{\gamma}_{i-\alpha}$ being the annihilation operator of a quasiparticle in the scattering state $i-\alpha$, with $i=u,d1,d2$ and $\alpha={\rm in,out}$. It is worth noting that the order of the creation/annihilation operators is changed for the $d2$ modes because of its anomalous character, as in this case the conjugate $\bar{z}_{{\rm {d2-\rm{in}},\omega}}$ is that with positive norm, $(\bar{z}_{{\rm {d2-\rm{in}},\omega}}|\bar{z}_{{\rm {d2-\rm{in}},\omega}})=1$. This is due to our choice of using only modes with positive energy to characterize the solutions of BdG equation; a complementary approach is considering only proper modes with positive norm, in which case the $d2$ modes have negative energy $\omega<0$. Nevertheless, both approaches give rise to the same relation as they are connected to each other through the symmetry $z\rightarrow\bar{z}$.

By considering the conservation of the quasiparticle current [Eqs. (\ref{eq:quasiparticlecurrent}), (\ref{eq:stationarycurrentssolutions})] for an arbitrary linear combination of ``in" scattering states, it can be shown that the $S$-matrix is pseudo-unitary, i.e., it satisfies
\begin{equation}\label{eq:pseudounitarity}
S^{\dagger}\eta S=\eta\equiv{\rm diag}(1,1,-1),
\end{equation}
so, in terms of group theory, $S\in U(2,1)$.

Finally, the quantum fluctuations of the field operator reads in terms of the scattering states as
\begin{eqnarray}\label{eq:BHFieldOperator}
\hat{\Phi}(x) & = & \int_{0}^{\infty}d\omega\sum_{a={\rm u-\rm{in},d1-\rm{in}}}[z_{a,\omega}(x)\hat{\gamma}_{a}(\omega)+\bar{z}_{a,\omega}(x)\hat{\gamma}_{a}^{\dag}(\omega)]\\
\nonumber &+&\int_{0}^{\omega_{{\rm max}}}d\omega[z_{{\rm {d2-\rm{in},\omega}}}(x)\hat{\gamma}_{{\rm {d2-\rm{in}}}}^{\dag}(\omega)\nonumber+\bar{z}_{{\rm {d2-\rm{in}},\omega}}(x)\hat{\gamma}_{{\rm {d2-\rm{in}}}}(\omega)] \, .
\end{eqnarray}
A similar expansion can be performed using the ``out'' scattering states, which at the end simply amounts to replacing ``in'' by ``out'' in the previous equation.

\subsubsection{Hawking effect}\label{subsec:hawkingeffect}

The key of the analogy with the Hawking effect arises when considering Eq. (\ref{eq:inoutmodesrelation}), as it is a Bogoliubov type relation that mixes creation and annihilation operators due to the anomalous character of the $d2$ modes. If one evaluates the expectation value of the number of outgoing $u$ phonons, $\braket{\hat{\gamma}_{u-\rm{out}}^{\dagger}\hat{\gamma}_{u-\rm{out}}}$, in the vacuum of the incoming modes, it is found that
\begin{eqnarray}\label{eq:hawkingef}
\braket{\hat{\gamma}_{u-\rm{out}}^{\dagger}(\omega)\hat{\gamma}_{u-\rm{out}}(\omega)}=|S_{ud2}(\omega)|^2\neq 0
\end{eqnarray}
This emission of outgoing quasi-particles into the subsonic region in the presence of the vacuum of incoming quasi-particles constitutes the spontaneous Hawking effect and it is a genuine quantum feature. It is the analog of the spontaneous emission of particles in a gravitational black hole, where the role of the outside of the black hole is played here by the subsonic region. As seen in Eq. (\ref{eq:hawkingef}), the intensity of the spontaneous Hawking signal is characterized by the scattering matrix element $S_{ud2}$.

A qualitative explanation of the Hawking effect can be obtained by examining the grand-canonical Hamiltonian $\hat{K}$. If we focus on the $\omega<\omega_{\rm max}$ sector of the Bogoliubov contribution [given by Eq. (\ref{eq:Bogoliubovhamiltonian})], we find
\begin{equation}\label{eq:BHK}
\sum_{i,j=u,d1,d2}\int_{0}^{\omega_{\rm max}}d\omega~\hbar\omega(\hat{\gamma}_{i-\rm{in}}^{\dagger}\eta_{ij}\hat{\gamma}_{j-\rm{in}})=\int_{0}^{\omega_{\rm max}}d\omega~\hbar\omega(\hat{\gamma}_{u-\rm{in}}^{\dagger}\hat{\gamma}_{u-\rm{in}}+\hat{\gamma}_{d1-\rm{in}}^{\dagger}\hat{\gamma}_{d1-\rm{in}}-\hat{\gamma}_{d2-\rm{in}}^{\dagger}\hat{\gamma}_{d2-\rm{in}}) \,.
\end{equation}

The minus sign of the $d2$ modes in the previous equation arises due to their negative norm. It can also be seen as the negative energy of the corresponding conjugate modes, conventionally normalized. Using Eqs. (\ref{eq:inoutmodesrelation}) and (\ref{eq:pseudounitarity}), Eq. (\ref{eq:BHK}) can be rewritten in terms of the ``out" states, which gives the same expression due to the pseudo-unitarity of the scattering matrix.

When working in black-hole analogues, one can essentially choose between two conventions, clearly represented in Fig. \ref{fig:DispRelation}: (i) All frequencies are taken as positive while the normalization of the scattering channels can be positive (plotted in blue) or negative (plotted in red) or (ii) All normalizations are taken as positive but then one has to deal with positive and negative frequencies. The first convention is more convenient to perform calculations so we have adopted it throughout this work, as quasi-particle scattering is viewed as elastic. The second convention provides, however, a simple physical picture of Hawking radiation: once there are positive and negative frequency outgoing scattering channels, one can expect that, even at zero temperature, two outgoing quasi-particles can be created spontaneously at zero-energy cost: the positive-frequency quasi-particle flows towards the subsonic side while the negative-energy partner flows towards the supersonic one. Within the first convention, the process of Hawking radiation can be viewed as the vacuum of incoming quasi-particles generating outgoing ``particle-antiparticle" pairs (involving channels $u$ and $d2$). Thus, the Hawking radiation corresponding to the emission of outgoing $u$-phonons into the subsonic region appears to an outside observer as spontaneously created by the horizon.

The conversion between two normal or two anomalous channels is labeled as a ``normal'' scattering process, while conversion from a normal to an anomalous channel (or vice versa) is labeled as an ``anomalous'' scattering process. Hawking radiation can be viewed as the result of anomalous scattering. Another anomalous process is the analog of Andreev reflection \cite{Zapata2009a}, similar to the Hawking effect but involving the $d1$ and $d2$ channels. Borrowing concepts from quantum optics, Hawking radiation can be understood as a non-degenerate parametric amplifier \cite{Walls2008}, as the vacuum of the incoming modes is a squeezed state for the outgoing modes.

\subsubsection{Computation of the scattering matrix}\label{subsec:SmatrixComputation}

Finally, we discuss the details of the computation of the $S$-matrix elements for a black-hole configuration. For that purpose, we study the matching of the BdG solutions in the asymptotic (subsonic and supersonic) regions. In 1D, as the stationary eigenvalue problem of the BdG equations (\ref{eq:BdGequation}) is a pair of second order differential equations in $x$ for the components $u(x),v(x)$ of the eigenmode $z(x)$, it can be rewritten as a first-order $4\times 4$ linear system:
\begin{equation}\label{eq:4BdG}
\frac{d\sigma}{dx}=D(x,\omega)\sigma,~\sigma(x)=\left[ u(x), v(x),u'(x),v'(x)\right]^{T}
\end{equation}
where $\hbar\omega$ is the energy of the mode and $D(x,\omega)$ is a $4\times4$ matrix of the form:
\begin{eqnarray}
D(x,\omega)&=&\left[\begin{array}{cc}
0 & \mathbb{I}_2\\
M(x,\omega) & 0
\end{array}\right]\\
\nonumber M(x,\omega)&=&\frac{2m}{\hbar^2}\left[\begin{array}{cc}
V(x)+2g|\Psi_0(x)|^2-\mu-\hbar\omega & g\Psi^2_0(x)\\
g\Psi^{*2}_0(x) & V(x)+2g|\Psi_0(x)|^2-\mu+\hbar\omega
\end{array}\right],
\end{eqnarray}
$\mathbb{I}_2$ being the identity matrix in two dimensions. In order to solve the problem, we divide the space in three regions: the subsonic region ($x\rightarrow-\infty$), the supersonic region ($x\rightarrow\infty$) and the scattering region (that between the subsonic and supersonic regions). In both asymptotic regions, the solutions are combinations of the different scattering channels, as explained before. By writing the matching conditions for $u,v$ and their derivatives at the points $x_u\rightarrow-\infty$ and $x_d\rightarrow\infty$ for $\omega<\omega_{\rm{max}}$, we find
\begin{eqnarray}\label{eq:udmatching}
\nonumber b_{u-\rm{in}}\sigma_{u-\rm{in},\omega}\left(x_u\right)&+&b_{u-\rm{out}}\sigma_{u-\rm{out},\omega}\left(x_u\right)
+b_{\rm{ev}}\sigma_{\rm{ev},\omega}\left(x_u\right)=\sum_{j=1}^{4}b_{j}\sigma_{i,\omega}\left(x_u\right)\\
\sum_{j=1}^{4}b_{j}\sigma_{i,\omega}\left(x_d\right)&=&b_{d2-\rm{in}}\sigma_{d2-\rm{in},\omega}\left(x_d\right)+b_{d1-\rm{in}}\sigma_{d1-\rm{in},\omega}\left(x_d\right)\\
\nonumber &+&b_{d1-\rm{out}}\sigma_{d1-\rm{out},\omega}\left(x_d\right)+b_{d2-\rm{out}}\sigma_{d2-\rm{out},\omega}\left(x_d\right)
\end{eqnarray}
with $\sigma_{a,\omega}$ a four-component vector that contains the spinor that describes the corresponding plane wave solution, $s_{a,\omega}(x)$, and its derivative,
\begin{equation}
\sigma_{a,\omega}(x)=\left[\begin{array}{c}
s_{a,\omega}(x)\\
s'_{a,\omega}(x)
\end{array}\right]
\end{equation}
The spinors $\sigma_{i,\omega},~i=1,2,3,4$ of Eq. (\ref{eq:udmatching}) represent four arbitrary independent solutions of the BdG equations in the scattering region while $\sigma_{\rm{ev},\omega}$ corresponds to the real exponential evanescent wave that exists in the subsonic region (the other exponential solution explodes and then it is unphysical). Thus, there are $11$ amplitudes and $8$ restrictions so we only have $3$ degrees of freedom. In particular, one can choose as independent the ``in" scattering channel amplitudes, in terms of which Eq. (\ref{eq:udmatching}) can be rewritten as a $8\times 8$ linear system of equations:
\begin{eqnarray}\label{eq:scatteringlinearsystem}
\nonumber d&=&A\cdot b\\
b&=&\left[b_{u-\rm{out}},b_{\rm{ev}},b_{1},b_{2},b_{3},b_{4},b_{d1-\rm{out}},b_{d2-\rm{out}}\right]^{T}\\
\nonumber d&=&b_{u-\rm{in}}\left[\begin{array}{c}\sigma_{u-\rm{in},\omega}\left(x_u\right)\\ 0 \end{array}\right]+b_{d1-\rm{in}}\left[\begin{array}{c}0\\ \sigma_{d1-\rm{in},\omega}\left(x_d\right)\end{array}\right]+b_{d2-\rm{in}}\left[\begin{array}{c}0\\ \sigma_{d2-\rm{in},\omega}\left(x_d\right)\end{array}\right],\\
\nonumber A&=&\left[\begin{array}{ccccc}-\sigma_{u-\rm{out},\omega}\left(x_u\right)&-\sigma_{\rm{ev},\omega}\left(x_u\right)&F(x_u)&0&0\\ 0&0&F(x_d)&-\sigma_{d1-\rm{out},\omega}\left(x_d\right)&-\sigma_{d2-\rm{out},\omega}\left(x_d\right)\end{array}\right]
\end{eqnarray}
The matrix $F(x)$ is the fundamental matrix in the scattering region, i.e., a $4\times 4$ matrix whose columns contain the four linearly independent solutions $\sigma_i(x)$.

In practice, the previous matching equations are solved by choosing the points $x_u,x_d$ sufficiently deep in the regions where the GP wave function tends to a plane wave. In typical configurations, the wave function tends to a plane wave with exponentially small corrections in the subsonic region and it is a perfect plane wave in the supersonic region for a finite value of $x_d$ \cite{Larre2012}. The fundamental matrix $F$, when cannot obtained analytically, is computed numerically by integrating the $4\times 4$ system of equations (\ref{eq:4BdG}) for four independent initial conditions. The fundamental matrices at two different points $x_1,x_2$ are connected through the transfer matrix $F(x_2)=T(x_2,x_1)F(x_1)$, which after inverting the relation gives
\begin{equation}\label{eq:transfermatrix}
T(x_2,x_1)=F(x_2)F^{-1}(x_1)
\end{equation}
In order to numerically compute $T(x_2,x_1)$, one can select the four independent solutions such that $F(x_1)=1$ and then the transfer matrix is simply $T(x_2,x_1)=F(x_2)$. An interesting property of the transfer matrix is that
\begin{equation}\label{eq:wronskian}
\det T(x_2,x_1)=1
\end{equation}
as the Wronskian $W(x)\equiv\det F(x)$ is a conserved quantity.

For theoretical purposes, we use Cramer's rule to formally solve the system of equations (\ref{eq:scatteringlinearsystem}):
\begin{equation}\label{eq:Cramer}
b_l=\frac{\textrm{det} ~A_l}{\textrm{det} ~A}, ~l=1,2...8 \, ,
\end{equation}
$b_l$ being the $l$ component of the vector $b$ and $A_l$ the matrix $A$ with the column $l$ replaced by the solution vector $d$. From here, it is straightforward to check that every amplitude is a linear function of the amplitudes of the ``in" modes. In particular, as the amplitude of the ``out" modes is related to the amplitude of the ``in" modes through the scattering matrix, the elements of the column $a$ of the scattering matrix are computed by setting the corresponding amplitude $b_{a-\rm{in}}=1$ and the remaining ``in" amplitudes to zero in the vector $d$ of Eq. (\ref{eq:scatteringlinearsystem}), and solving the resulting system. However, we note that, in the actual computation, Eq. (\ref{eq:scatteringlinearsystem}) is solved numerically.

\section{Numerical integration of the quasi-stationary BdG equations}\label{app:numerical}

In order to obtain the scattering matrix in the quasi-stationary regime, we need to solve the system of equations (\ref{eq:scatteringlinearsystem}). The numerical computation of the fundamental matrix $F(x)$ presents some problems in the optical lattice as one of the eigenvalues of the fundamental matrix grows exponentially, corresponding to a local Bloch wave with complex wave vector. For sufficiently large optical lattices, this makes that the fundamental matrix becomes singular within computer's relative accuracy, spoiling the calculation. In order to deal with this problem, we consider two methods: QR decomposition \cite{Slevin2004} and the Global Matrix method \cite{Lowe1995}. Both methods are based on the following property of the transfer matrix of Eq. (\ref{eq:transfermatrix}):
\begin{equation}\label{eq:transferdecom}
T(x_{n+1},x_1)=\prod_{i=1}^{n}T_i,~T_i\equiv T(x_{i+1},x_{i})
\end{equation}
with $x_1<x_2<\ldots<x_{n+1}$ being arbitrary intermediate points, i.e., the total transfer matrix can be decomposed as the product of intermediate transfer matrices. In our problem, we have to compute the transfer matrix $T(x_d,x_u)$ between the two asymptotic regions so $x_1=x_u$ and $x_{n+1}=x_d$. The two techniques here considered try to obtain the previous matrix without directly evaluating the previous multiplication as this would give a numerically singular matrix. The matrices $T_i$ are obtained by integrating the effective BdG equations (\ref{eq:effectiveTIBdG}) on top of the quasi-stationary GP wave function between the intermediate points $x_{i}$ and $x_{i+1}$, taken to be sufficiently close one to each other so all the eigenvalues of $T_i$ are of order $1$. Specifically, we use a Runge-Kutta method of 4th-order to integrate the BdG equations, choosing as numerical grid that of the quasi-stationary GP wave function, computed following the scheme described in Ref. \cite{deNova2014a}.

We have checked that both methods work correctly by comparing the results with the brute force result, in which the total transfer matrix is computed by direct multiplication of Eq. (\ref{eq:transferdecom}) after increasing the accuracy to 1000 digits. Of course, brute force represents a much slower method, motivating the use of these techniques. In order to further check the validity of the numerical results, we have computed the magnitude
\begin{equation}\label{eq:pseudounitarynorm}
\delta(\omega)\equiv S^{\dagger}(\omega)\eta S(\omega)-\eta
\end{equation}
which should be zero for a perfect pseudo-unitary matrix satisfying Eq. (\ref{eq:pseudounitarity}). We have found that $\delta(\omega)$ is much smaller than the transmission coefficients computed from the $S$-matrix, showing the reliability of the plots. Finally, we wish to note that there are no significant differences in the computational speed between the techniques so they both represent good choices to do the job.

\subsection{QR decomposition}

As well known from standard results in linear algebra, a given matrix $M$ can be decomposed as $M=QR$, with $Q$ orthogonal and $R$ an upper-diagonal matrix in the so-called the QR decomposition. The point is that one can compute the QR decomposition of the total transfer matrix by performing the QR decomposition at every step of the product of Eq. (\ref{eq:transferdecom}). Assuming that $T(x_{k},x_{1})=Q_kR_k$, with $k<n+1$, then $T(x_{k+1},x_{1})=T_{k+1}T(x_{k},x_{1})=T_{k+1}Q_kR_k$ and the QR decomposition of $T(x_{k+1},x_{1})=Q_{k+1}R_{k+1}$ results from the relation
\begin{equation}
T_{k+1}Q_k=\tilde{Q}_k\tilde{R}_k,
\end{equation}
which gives
\begin{equation}
Q_{k+1}=\tilde{Q}_k,~R_{k+1}=\tilde{R}_kR_{k}.
\end{equation}
Hence, the $Q,R$ matrices of the global transfer matrix, $T(x_d,x_u)=QR$, are obtained by applying this recursive process from $k=2$ to $k=n$.

Once $Q,R$ are obtained, the fundamental matrix in the supersonic region is related to that of the subsonic region through
\begin{equation}
F(x_d)=QRF(x_u)
\end{equation}
The point is that if one directly multiplies $QR$ in this equation, a singular matrix is obtained within computer's accuracy. Instead of that, we first diagonalize the $R$ matrix, $R=MDM^{-1}$, with $D$ a diagonal matrix; this diagonalization does not carry any associated numerical problem as $R$ is an upper-diagonal matrix. Since $Q$ is orthogonal, $\det Q=1$ and then also $\det R=\det D=\det T(x_d,x_u)=1$. In particular, the matrix $D$ presents two eigenvalues exponentially large (small), denoted as $\lambda_{+,-}$ (arising from the Bloch solutions with complex wave vector) and other two eigenvalues of order $\sim 1$ (associated to the propagating Bloch waves). The numerical problem arises from mixing the columns of $D$, since $\lambda_{+}$ crushes the eigenvalues of order $1$ as they are so relatively small that they fall under computer's relative accuracy.

For fixing this numerical issue, we first rearrange $D$ so the column corresponding to $\lambda_+$ is the first one. We then choose $F(x_u)$ as $F(x_u)=M I_+$, where $I_+$ is the $4\times 4$ identity but with the first diagonal element replaced by $\lambda^{-1}_+$. As a consequence, we obtain a fundamental matrix in the supersonic region $F(x_d)=QM\tilde{D}$, with $\tilde{D}$ the same matrix as $D$ but with the exponentially large element $\lambda_+$ replaced by $1$. In this way, the problematic columns of $F(x_{u,d})$ are not mixed and the Wronskian is conserved, suppressing any possible singular behavior. We note that for extremely long optical lattices a problem of overflow can appear due to the exponentially large value of $\lambda_+$. However, that extreme case is not a problem since then one can always make use of the Global Matrix method.

\subsection{Global Matrix method}

The previous method just rearranges the product of Eq. (\ref{eq:transferdecom}) in a clever way in order to obtain a non-singular matrix. The method here presented avoids the explicit computation of the total transfer matrix. Instead of that, the Global Matrix method rewrites Eq. (\ref{eq:scatteringlinearsystem}) as a $(4n+4)\times(4n+4)$ system of equations by considering the matching of the solutions at every intermediate point $x=x_{i}$, $i=2,3,\ldots,n$

\begin{eqnarray}\label{eq:GlobalMatrix}
\nonumber d&=&A\cdot b\\
b&=&\left[b_{u-\rm{out}},b_{\rm{ev}},b_{1},b_{2},\ldots,b_{4n-1},b_{4n},b_{d1-\rm{out}},b_{d2-\rm{out}}\right]^{T}\\
\nonumber d&=&b_{u-\rm{in}}\left[\begin{array}{c}\sigma_{u-\rm{in},\omega}\left(x_u\right)\\ 0\\ \vdots \\ 0  \end{array}\right]+b_{d1-\rm{in}}\left[\begin{array}{c}0\\ \vdots \\ 0 \\   \sigma_{d1-\rm{in},\omega}\left(x_d\right)\end{array}\right]+b_{d2-\rm{in}}\left[\begin{array}{c}0\\ \vdots \\ 0 \\ \sigma_{d2-\rm{in},\omega}\left(x_d\right)\end{array}\right],\\
\nonumber A&=&\left[\begin{array}{cccccccc}-\sigma_{u-\rm{out},\omega}&-\sigma_{\rm{ev},\omega}&\mathbb{I}_4&0&0&\ldots&0&0\\
0&0&T_1&-\mathbb{I}_4&0&\ldots&0&0\\
0&0&0&T_2&-\mathbb{I}_4&0&\ldots&0\\
~&~&~&~&\ddots&\ddots&\ddots&~\\
0&0&0&0&T_{n-1}&-\mathbb{I}_4&0&0\\
0&0&0&0&0&T_n&-\sigma_{d1-\rm{out},\omega}&-\sigma_{d2-\rm{out},\omega}\end{array}\right]
\end{eqnarray}
where $\mathbb{I}_4$ is the $4\times 4$ identity matrix and the spatial dependence of the vectors $\sigma$ inside the matrix $A$ is understood for simplicity. Hence, the numerically singular behavior of the total transfer matrix is avoided by solving at one time the {\it global} problem.

\bibliographystyle{apsrev4-1}
\bibliography{Hawking}

\end{document}